\documentclass[fleqn,usenatbib]{mnras}

\usepackage{newtxtext,newtxmath}

\usepackage[T1]{fontenc}
\usepackage{comment}

\DeclareRobustCommand{\VAN}[3]{#2}
\let\VANthebibliography\thebibliography
\def\thebibliography{\DeclareRobustCommand{\VAN}[3]{##3}\VANthebibliography}

\usepackage{graphicx}	
\usepackage{amsmath}	
\usepackage{tikz}
\usepackage{algpseudocode}
\usetikzlibrary{arrows.meta, positioning, shapes, calc}

\title[Adaptive Fast Multipole Method in RAMSES]{A Scalable Fast Multipole Method Poisson Solver for the RAMSES code: II. Adaptive Mesh Refinement and Adaptive Time Stepping}

\author[J. -Y. Lee and R. Teyssier]{
Jun-Young Lee,$^{1}$\thanks{E-mail: junyoung.lee@princeton.edu}
Romain Teyssier,$^{1}$
\\
$^{1}$Department of Astrophysical Sciences, Princeton University, 4 Ivy Lane, Princeton, NJ 08544, USA\\
}

\date{Accepted XXX. Received YYY; in original form ZZZ}

\pubyear{\the\year{}}

\begin{document}
\label{firstpage}
\pagerange{\pageref{firstpage}--\pageref{lastpage}}
\maketitle

\begin{abstract}
We present an extended implementation of a scalable, $\mathcal{O}(N)$ Poisson solver based on the fast multipole method (FMM), fully compatible with adaptive mesh refinement (AMR) and adaptive time stepping (ATS) within the RAMSES framework. Building on the unigrid algorithm in \citet{FMM_Unigrid_RAMSES}, we introduce several novel elements, including the use of multiple FMM trees, one for each AMR level, a merged FMM tree for all active levels to optimize neighbor searches, and the introduction of the concept of ``nearest-field'' to enforce force symmetry across refinement levels. Across a broad set of test problems, we find excellent, percent-level agreement with our reference traditional multigrid (MG) solver. We show, however, that FMM exhibits better momentum conservation properties across coarse-fine AMR interfaces. Finally, despite the overhead introduced by the spatio-temporal adaptivity, FMM shows better scalability than MG across various AMR configurations, with the largest gains obtained for the largest configurations.
\end{abstract}




\section{Introduction}

Understanding the evolution of self-gravitating astrophysical fluids requires resolving multi-scale physics, especially as gravity and complex baryonic physics interplay. This presents a major numerical challenge because accurately describing small-scale astrophysical processes demands high spatial and temporal resolution, which can substantially constrain the global timestep and overall computational cost. A key breakthrough is adaptivity in both space and time, enabling highly refined regions with shorter characteristic timescales to be evolved locally, while avoiding unnecessary cost in smoother, coarser regions. Astrophysical codes achieve this adaptivity in different ways, particularly in their treatment of hydrodynamics and feedback. 

In Lagrangian particle-based methods such as Smoothed Particle Hydrodynamics \citep[SPH;][]{Gingold_Monaghan_1977}, resolution naturally follows the mass through adaptive smoothing lengths, whereas in Eulerian grid-based methods, resolution is increased through adaptive mesh refinement \citep[AMR;][]{Berger_Oliger_1984,BergerColella}. Different resolution elements are evolved through individual time steps, usually limited by the shortest local physical time scale that govern the system. In addition, diverse refinement criteria can be applied to closely follow the local physics while keeping the numerical simulations within realistic run times. For example, cosmological simulations solving gravity typically employ refinement criteria based on the masses or the Jeans length of a given cell \citep{Truelove1997}.

Unlike hydrodynamics, which is roughly local as information travels at finite sound speed and is limited by the Courant-Friedrich-Levy conditions for numerical stability, gravity is governed by the Poisson equation, which is an elliptic PDE, and its solution depends on the global, non-local density field. It is therefore important, but also challenging, to communicate this global information consistently across multi-resolution grids, where the data structure is irregular and numerical errors can arise near refinement boundaries. In addition, the solver scaling can depend strongly on the bespoke implementation choices made to accommodate the irregular data structures introduced by refinement and the detailed number of arithmetic operations, in addition to the nominal algorithmic complexity of the solver.

The Fast Fourier Transform (FFT), for example, has $\mathcal{O}(N\log N)$ complexity, but is highly efficient on uniform Cartesian grids and can outperform many nominally $\mathcal{O}(N)$ methods \citep{Hockney_Eastwood_1988}. However, FFT-based solvers are less well suited to highly refined AMR grid structures. In Eulerian AMR codes, multigrid \citep[MG;][]{BAKHVALOV1966101, FEDORENKO19621092, Brandt1977MultilevelAS}, which have $\mathcal{O}(N)$ complexity, are therefore widely adopted. In MG, a hierarchy of auxiliary grids is constructed on top of the AMR hierarchy, and iterative relaxation, restriction, and prolongation are used across levels to accelerate convergence on large spatial scales. MG is computationally lightweight, since each relaxation sweep involves only a small stencil operation, and is memory efficient, as the auxiliary grids can be built sparsely around the patchy AMR regions \citep[e.g.,][]{GUILLET_2011}. However, MG typically requires multiple V-cycles to reach convergence and has low arithmetic intensity, meaning that it performs relatively few floating-point operations (FLOPs) per data movement. This leaves room for alternative methods that may better exploit modern hardware that is well-suited for compute-intensive algorithms.

To address this issue, in \citet{FMM_Unigrid_RAMSES} (hereafter Paper I), we have recently explored an alternative algorithm with linear time complexity, the Fast Multipole Method \citep[FMM;][]{Greengard_Rokhlin_1987,Greengard_Rokhlin_laplace_3d,1997AcNum...6..229G}, in the unigrid setting of RAMSES  \citep{RAMSES_2002}. In contrast to MG, FMM sweeps the multi-resolution grid only once---upward sweep for multipole accumulation, and downward sweep for local expansions and force evaluations---and can benefit from less frequent neighborhood searches entailing data movement and MPI communication overhead. Especially, our unigrid FMM implementation revealed that although the number of FLOPs is substantially larger than that of MG, exceeding it by more than a factor of 30, the FMM algorithm exhibits comparable performance and better scaling behavior.

FMM has been widely used in particle-based Lagrangian codes and $N$-body codes \citep{Dehnen2014_stellar_dynamics,GADGET-4,PKDGRAV3,SWIFT}. In many FMM particle-based implementations, spatial adaptivity is typically controlled by opening-angle criteria (or the multipole acceptance criteria), as in tree-based gravity solvers. These criteria determine whether an interaction should be approximated by cell-cell or cell-particle interaction rather than treated through direct particle-particle interaction. 

On the other hand, many Eulerian codes that introduce spatial adaptivity through AMR solve Poisson's equation by discretizing the Laplacian with finite differences and solving the resulting matrix equation using iterative methods, usually variants of MG, occasionally combined with FFTs for the long-range periodic component along the mesh structure \citep{art,RAMSES_2002,Castro,Enzo,GamerI}. FMM-based alternatives are comparatively less common in this setting, although they have been explored recently. For example, \citet{OctoTiger} uses multipoles constructed on an oct-tree AMR hierarchy, while employing an opening-angle criterion similar to those used in particle-based codes.

More generally, adaptive versions of FMM have also been developed, in which the tree is refined until each leaf cell contains no more than a prescribed number of particles \citep{Hrycak_Rokhlin1998,Cheng_Greengard_Rokhlin_1999,Ethridge&Greengrad2001}. However, these methods are formulated primarily for particle-based problems on a single globally synchronized and spatially adaptive tree, with interaction lists determined by tree geometry and particle-counts. They are also not directly suited to the Adaptive Time Stepping (ATS) scheme in RAMSES, where only the fine active levels are advanced while coarser levels remain inactive, requiring active level multipoles to be refreshed without discarding the stale contributions from inactive levels.

We present an algorithm that follows both the spatial adaptivity of the oct-tree AMR hierarchy and the temporal adaptivity introduced by sub-cycling. In contrast to previous FMM implementations that operate directly on particles, our implementation of FMM in RAMSES is combined with the Particle-Mesh (PM) method. In the adaptive version of the PM scheme, the particle masses are deposited onto the AMR cells using standard mass deposition schemes such as CIC or TSC. The FMM interaction geometry is therefore inherited directly from the native AMR grid hierarchy, rather than being determined by conventional opening-angle criteria. Since RAMSES adopts cell-by-cell refinement, the AMR grid can efficiently follow complex geometries without introducing unnecessary refinement over larger blocks. Moreover, because the refinement criteria can be adapted to the physical system of interest, the FMM geometry tied directly to the AMR geometry and can thus naturally inherit the same problem-dependent adaptivity.

The fine-grained, cell-based refinement strategy of RAMSES introduces algorithmic challenges, since it leads to a more irregular data structure than block-based AMR schemes. At the same time, the AMR hierarchy has useful structural properties: the mesh is graded, so adjacent cells differ by at most one refinement level, and efficient neighbor finding is supported by fully threaded trees or fast hash tables. A central challenge of our implementation is therefore to adapt FMM to this highly flexible AMR structure in RAMSES while retaining efficiency and scalability.

Moreover, we also incorporate the ATS algorithm, realized by sub-cycling finer levels in finer time steps. In ATS, coarser levels are frozen in time and deemed inactive, while all finer levels are advanced in synchronized steps and deemed active. This complicates the FMM algorithm, as the multipoles from the active levels would need to be refreshed, while we still need to store the stale multipoles from the inactive levels. This requires storing separate FMM hierarchies per refinement level and identifying neighbors across these different hierarchies. 

To address this, we first introduce the concept of multiple sparse FMM trees, which are maintained by each of the refinement levels. To speed up the repeated neighborhood searches and interactions, we introduce the new concept of \textit{merged tree} (Section~\ref{sec:trees}). Additionally, to ensure force symmetry across different refinement levels, we also introduce the new concept of the \textit{nearest field} (Section~\ref{sec:nearest}), in addition to the traditional FMM concepts of far, intermediate and near fields. The life cycle of multiple FMM trees for the ATS and sub-cycling along with the full overview of the algorithm is introduced in Section \ref{sec:lifecycle}. Section \ref{sec:test} compares MG and FMM in different test cases, and Section~\ref{sec:scaling} demonstrates the scaling behavior of our code. Finally, Section~\ref{sec:conclusion} summarizes the algorithm and results.

\section{The Unigrid Algorithm}
FMM is an algorithm with linear $\mathcal{O}(N)$ complexity, which is realized by approximation of long-range forces through multipole expansions instead of pairwise interactions and approximation of the local gravitation field from Taylor expansions. Usually, a tree or a multi-resolution grid is created initially, onto which the multipoles are accumulated and shifted. Following Paper I, we construct FMM grids on top of the AMR grid and accumulate the cells' masses $m_{\rm cell}$ to multipoles $\mathcal{M}_n$ up to the quadrupole order ($n=2$). As RAMSES organizes $2^d$ cells into a single oct in $d$-dimensions, the finest FMM oct comprises $2^d$ octs of the AMR grid\footnote{Paper I tested a generalized case where the level difference between the AMR grid and the coarsest FMM grid can be arbitrary by $\Delta \ell$, which controls the size of the near field region at the bottom level of the downward pass. However, while the accuracy was nearly identical for both $\Delta \ell=1$ and 2, the enlarged kernel for direct force calculation worsened the scaling performance. Also, the given complexity of the algorithm given $\Delta\ell=2$, we fix $\Delta\ell=1$ for the AMR and ATS algorithm.}. Thus, the multipoles are accumulated by the following particle-to-multipole operation (P2M)\footnote{For consistency with the conventional FMM terminology and with Paper I, we refer to the operations using the standard particle-based names such as P2M, M2P, and L2P. In our implementation, however, the fundamental source and sink elements are not particles but grid cells, onto which particle masses have already been deposited. Thus, what would conventionally be called P2M should be understood here as a cell-to-multipole operation, i.e. the construction of multipole moments from the mass distribution defined on grid cells.}:
\begin{eqnarray} \label{eqn:multipole}
    \mathcal{M}_n = \sum_{\rm cell_i \in oct} m_{\rm cell} \,  {\bf x_i}^{(n)},
\end{eqnarray}
where ${\bf x_i}$ is the cell's center and $(n)$ is the $n$-fold outer product. The multipoles for the coarser FMM cells are accumulated from the finer FMM cells after being shifted to the center of the coarser cell, following multipole-to-multipole (M2M) operation:
\begin{eqnarray}
\label{eq:M2M_general}
\mathcal{M}'_n =
\sum_{k=0}^{n}
\binom{n}{k}
(-\mathbf{a})^{(n-k)}
\odot
\mathcal{M}_k,
\end{eqnarray}
where $\mathbf{a}$ is the displacement and $\odot$ a tensor inner product. The M2M operation is performed at each FMM level up to the coarsest FMM grid, and after the upward pass, each FMM cell will hold its multipole information.

A deliberate design choice was made to expand the multipoles with respect to the cell centers, using a Cartesian basis to compute the distance kernels beforehand and reduce the number of operations on-the-fly. Although we used a limited number of multipoles, we achieved precision comparable to that of MG. Although higher polynomial degrees could be possible for better precision, this would correspond to the MG algorithm using a higher-order finite difference approximation of the Laplacian operator, which is out of the scope of our paper. Using a relatively low polynomial order enables lean memory usage, which is critical for our future optimization of GPU architectures to reduce memory pressure. 

After the multipoles are accumulated through the upward pass along the multi-resolution grid hierarchy, the force is evaluated during the downward pass toward the finest FMM level. For each target FMM cell at a given level, the source domain is divided into three parts: the far-field, the intermediate-field, and the near-field. The intermediate-field is given by a set of well-separated FMM cells, which forms the target cell's interaction list. In a $d$-dimensional unigrid, this list contains $6^d - 3^d$ cells: the cells surrounding the father oct, excluding the $3^d$ cells that are in direct contact with the father cell. These excluded cells form the near field at the current level, and their contribution is instead accounted for through the intermediate-field calculation at the next finer level. The far-field corresponds to the remaining domain, whose contribution has already been included through the intermediate-field calculation of the coarser FMM cell. 

In short, the contribution from the intermediate field at each level is propagated down the hierarchy as a local expansion, represented by the Taylor coefficients of the gravitational field. For a given target cell, these coefficients are computed from the multipoles of well-separated source cells through multipole-to-local (M2L) operations:
\begin{eqnarray}
\label{eq:M2L}
\mathcal{L}^{p}_m
=
\sum_{n=0}^{p-m}
\frac{(-1)^n}{n!}
\nabla^{(n+m)} g(\mathbf{R})
\odot
\mathcal{M}_n ,
\end{eqnarray}
where $p$ is the polynomial order of the local expansion, $\mathcal{M}_n$ is the multipole of order $n$, and $g(\mathbf{R})$ is the Green's function of the Poisson's equation given a displacement of $\mathbf{R}$. Following Paper I, we use $p=3$, and the cell-centered scheme enables the precalculation of $\nabla^{(n+m)} g(\mathbf{R})$ tensors, reducing the arithmetic cost. 

The local expansions are approximations of the potential field originating from the far-field. Since the sources are sufficiently remote, the local expansions can be shifted from the parent cell's center to the target cell's center using the following local-to-local (L2L) operation:
\begin{eqnarray}
\label{eq:L2L}
{\mathcal{L}^{p}_m}^\prime
=
\sum_{n=0}^{p-m}
\frac{1}{n!}
\mathbf{a}^{(n)}
\odot
\mathcal{L}^{p}_{m+n},
\end{eqnarray}
where $\mathbf{a}$ is the displacement and the primed being the value after the translation. The newly computed local expansion from the M2L operations at the current level is added to the translated far-field local expansion, and the resulting expansion is passed to the next FMM call at the finer level. Once the downward pass reaches the finest FMM level, the nearest-field contribution is evaluated through pairwise $1/r$ interactions, often referred to as particle-to-particle (P2P) operations. The FMM algorithm is closely related to the Barnes Hut tree algorithm \citep{Barnes_Hut_1986}, which has $\mathcal{O}(N\log N)$ complexity, but achieves linear complexity by reusing local expansion terms among nearby target cells.

\begin{figure}
	\includegraphics[width=\columnwidth]{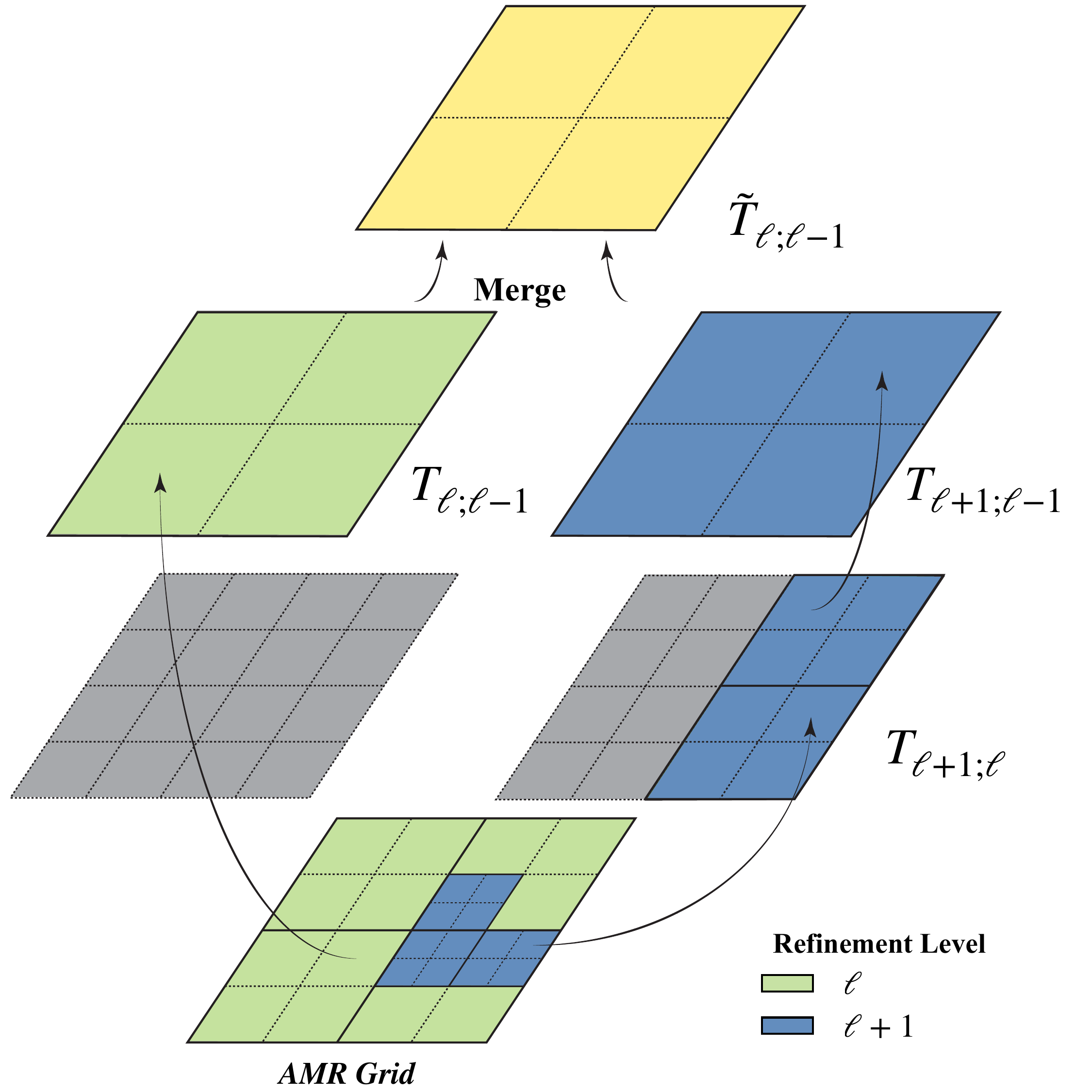}
    \caption{Visualization of the construction of a typical FMM tree over the chosen region. The \textit{green} and \textit{blue} cells in the AMR grid correspond to refinement levels $\ell$ and $\ell+1$, respectively. A grid $\mathcal{T}_{\ell; \ell'}$ is defined as a tree that starts from cells with refinement level $\ell$ and consists of FMM cells at level $\ell'$, with the cells colored according to the refinement level from which they originate. Gray areas mark locations where an FMM oct is unnecessary because no AMR cell is encompassed by the tree. The \textit{yellow} grid $\tilde{\mathcal{T}}_{\ell, \ell-1}$ represents the merger of the existing trees at a specified FMM across the active levels, and is created at every possible FMM tree level.}
    \label{fig:tree}
\end{figure}

\begin{figure*}
	\includegraphics[width=0.8\textwidth]{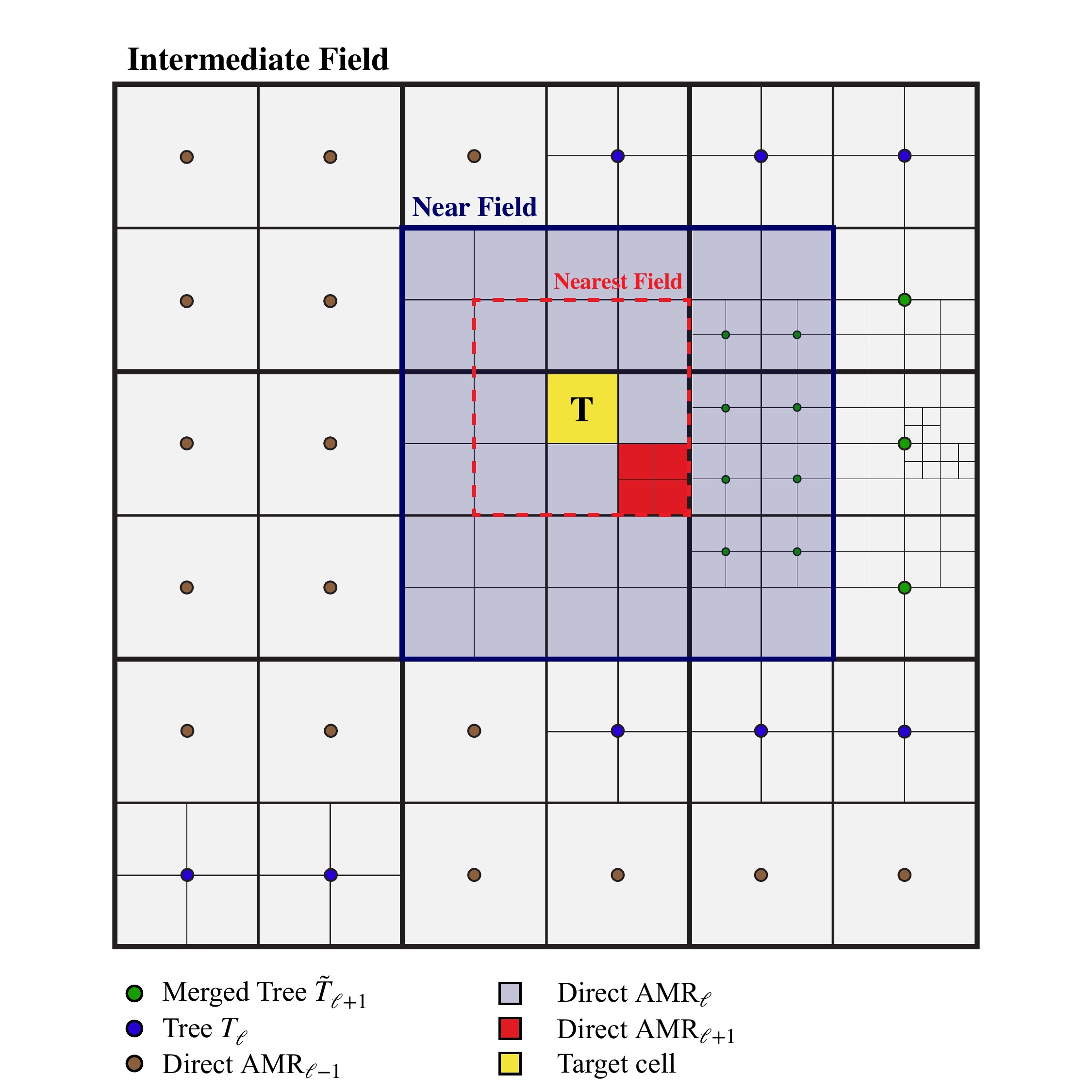}
    \caption{Anatomy of the force-field on an AMR grid example, featuring the intermediate field, near field, and the nearest field, centered around the target cell with refinement level of $\ell$ colored in \textit{yellow}. Firstly, the intermediate field is calculated from \textit{(i)} \textit{gray} AMR cells at level $\ell-1$ using P2P (\textit{brown dots}), M2P from either \textit{(ii)} $\mathcal{T}_\ell$ (\textit{blue dots}) or the \textit{(iii)} merged tree $\tilde{\mathcal{T}}_{\ell+1}$ (\textit{green dots}) at FMM level $\ell-1$. Secondly, the near field is calculate either from \textit{(i)} P2P with \textit{blue} AMR cells at level $\ell$ or \textit{(ii)} M2P from the merged tree $\tilde{\mathcal{T}}_{\ell+1}$ (\textit{green dots}). However, an exception is made for the cells in direct contact with the target cell, in which if the cells are further refined to $\ell+1$, we perform a P2P from each of the cells instead of M2P from FMM tree level of $\ell$ as done for the cells inside the near field but not in the nearest field.}
    \label{fig:field}
\end{figure*}

\begin{figure*}
	\includegraphics[width=\textwidth]{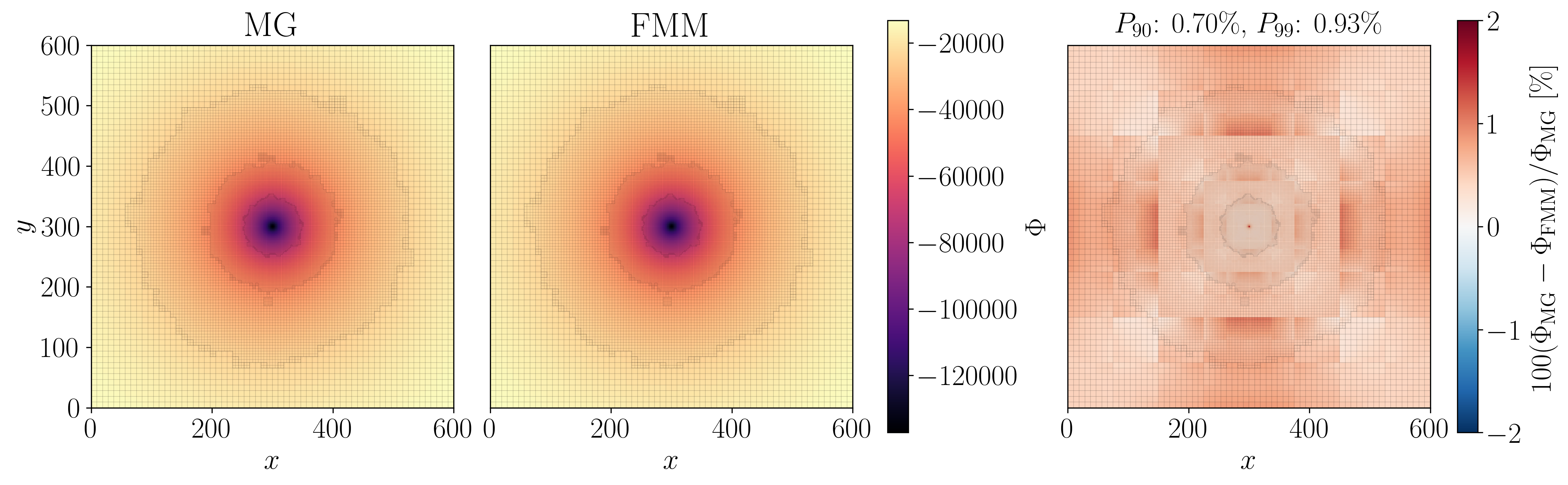}
    \caption{Comparison of the isolated NFW halo's graviational potential map solved on a four-level AMR grid between MG and FMM. The \textit{rightmost panel} exhibits that both the 90th- and 99-th percentiles of the mean relative difference are sub-percent. Apart from the largest source of error at the central peak, the remaining error is mostly concentrated near the coarse FMM cell boundaries. The different AMR cell sizes are also visible in the potential maps.}
    \label{fig:nfw}
\end{figure*}

\section{The Adaptive Extension}
\label{sec:adaptive}

\subsection{Multiple FMM Trees} \label{sec:trees}
In our unigrid implementation of FMM, we have constructed a single FMM grid hierarchy (or \textit{FMM tree}) from which the multipole accumulation and local expansion calculation were performed. However, for ATS we need to sub-cycle levels with higher refinement, which requires a separate treatment for each refinement level. Therefore, we construct distinct FMM trees per refinement level, which carry their own multipoles and local expansions. For a tree built on the refinement level $\ell$, we express it as $\mathcal{T}_\ell$. Here, the value of $\ell$ is drawn from $[\ell_{\rm min}, \ell_{\rm max}]$. Again, the FMM tree is a hierarchical multi-resolution grid and $\mathcal{T}_{\ell;\ell'}$ denotes the FMM grid at level $\ell'$, where $1\le\ell'\le\ell-1$\footnote{Efficiently allocating memory \textit{ab initio} for storing the FMM trees is non-trivial, because the AMR structure evolves dynamically and the required array size can change simultaneously across levels. Moreover, for deeper AMR hierarchies, the FMM implementation requires more memory than MG, since duplicate FMM cells can exist at the same position and size while carrying different multipole information. However, the prototype code \texttt{mini-ramses} remains relatively lean in memory usage, and the maximum overhead for maintainig the multi-resolution grid is only about $N/7$ per level. Worse case scenario, users can always restart the run if the initially allocated memory becomes insufficient.}.

Figure~\ref{fig:tree} illustrates how standard FMM trees are constructed from an AMR grid. The AMR grid contains level-$\ell$ cells in green and level-$(\ell+1)$ cells in blue. As shown in $\mathcal{T}_{\ell;\ell-1}$, the FMM tree does not span the entire computational domain but is built instead in the minimal region required to cover the refined area. The upward pass is carried out analogously to the unigrid case by adding only the multipoles associated with its own refinement level. For green AMR cells, the tree is constructed solely starting from the level of the $\ell-1$ FMM tree. The FMM grids $\mathcal{T}_{\ell;\ell-1}$ and $\mathcal{T}_{\ell+1;\ell-1}$ cover the same spatial domain but store different multipole data, since each FMM tree accumulates multipoles only from its corresponding AMR cells. At the top, we demonstrate an FMM grid of a merged tree $\tilde{\mathcal{T}}_{\ell;\ell-1}$, which aggregates the multipoles stored from all FMM grids at level $l-1$, or $\mathcal{T}_{\ell';\ell-1}$ with any possible value of $\ell'$. The merged tree is useful for optimization of the algorithm and will be explained later. 

\subsection{Extended Force Field: the Nearest Field}\label{sec:nearest}
In our unigrid FMM implementation, the force field is divided into three groups: far, intermediate, and near. The intermediate field is a collection of well-separated cells of $6^d-3^d$ in dimension $d$. The far field is a collection of cells further than the intermediate field, and its contribution is calculated in the previous step. The near field is the $3^d$ FMM cells in direct contact with the target FMM cell. The recursive nature of FMM naturally originates from the near field being the intermediate field of the following step. When the bottom of the hierarchy is reached, the near field is calculated via a direct force summation. 

For the AMR version of the FMM algorithm, the interaction geometry is inherited directly from the RAMSES mesh, rather than from a particle-count adaptive tree or an opening-angle criterion. We therefore introduce an extension to the force groups, the \textit{nearest field}. The nearest field is defined at the bottom of the hierarchy as the region in direct contact with the cell of interest ($3^d$ AMR cells including itself). If any of these neighboring cells are further refined, we calculate the direct force from the finer cells. This treatment exactly preserves force symmetry across coarse-fine interfaces\footnote{For particle-based codes employing a dual-tree walk with a symmetric multipole acceptance criteria, such a further decomposition of the force field is unnecessary, since pairwise interactions are evaluated symmetrically and the resulting forces are momentum-conserving \citep[e.g.,][]{Dehnen2000_symmetry}.}.

Figure~\ref{fig:field} shows the nearest field given a yellow target cell at the level $\ell$. In contrast to the near field shaded in blue, the nearest field is in direct contact with the AMR cell, instead of the AMR oct. Depending on the refinement geometry, if there are cells further refined as colored in red, we use the finer cells as the source. Again, this is to preserve the symmetry of forces as red cells will become the target and yellow the source when evaluating FMM at level $\ell+1$. The rigid graded-octree refinement strategy of RAMSES only allows a cell to be further refined if $3^d$ neighboring father cells are refined. Thus, the nearest field can consist only of cells with a refinement level of $\ell$ or $\ell+1$ at most, which simplifies the algorithm. 

The same AMR refinement rule also restricts the finest cells within the near field of the target to be at most one level deeper. However, the intermediate field is much more complex. As seen from Figure~\ref{fig:field}, the intermediate field can consist of varying refinement levels. The standard M2P operation is done from the FMM grid $\mathcal{T}_{\ell;\ell-1}$ as denoted in blue dots representing their multipoles. Given the refinement rule, there is a lower bound of level $\ell-1$ for AMR cells within the intermediate and near field. For these cells, the FMM tree starts at grid level $\ell-2$, and thus we would have to resort to a direct P2P force sourced from the brown dots, which holds only the mass information of the cell. In contrast to the coarser bound, the intermediate field can be refined deeply. In this scenario, we do the usual M2P operation sourced from the the FMM grid of $\mathcal{T}_{\ell';\ell-1}$, where we iterate over the possible values of $\ell'$. However, further optimization can be performed by merging the trees into $\tilde{\mathcal{T}}_{\ell+1}$, which aggregates all the multipoles of the trees $\mathcal{T}_{\ell'\ge\ell+1}$. This reduces the cost of multiple iterations over neighbors into a single intermediate force calculation from the FMM grid 
$\tilde{\mathcal{T}}_{\ell+1;\ell-1}$. The merged multipoles are highlighted in green dots. Similarly for $\ell+1$ cells in the near field, we can use M2P from $\tilde{\mathcal{T}}_{\ell+1;\ell}$. Unlike $\ell+1$ cells within the nearest field, the contribution is calculated from the M2P operation.

\begin{figure*}
	\includegraphics[width=0.7\textwidth]{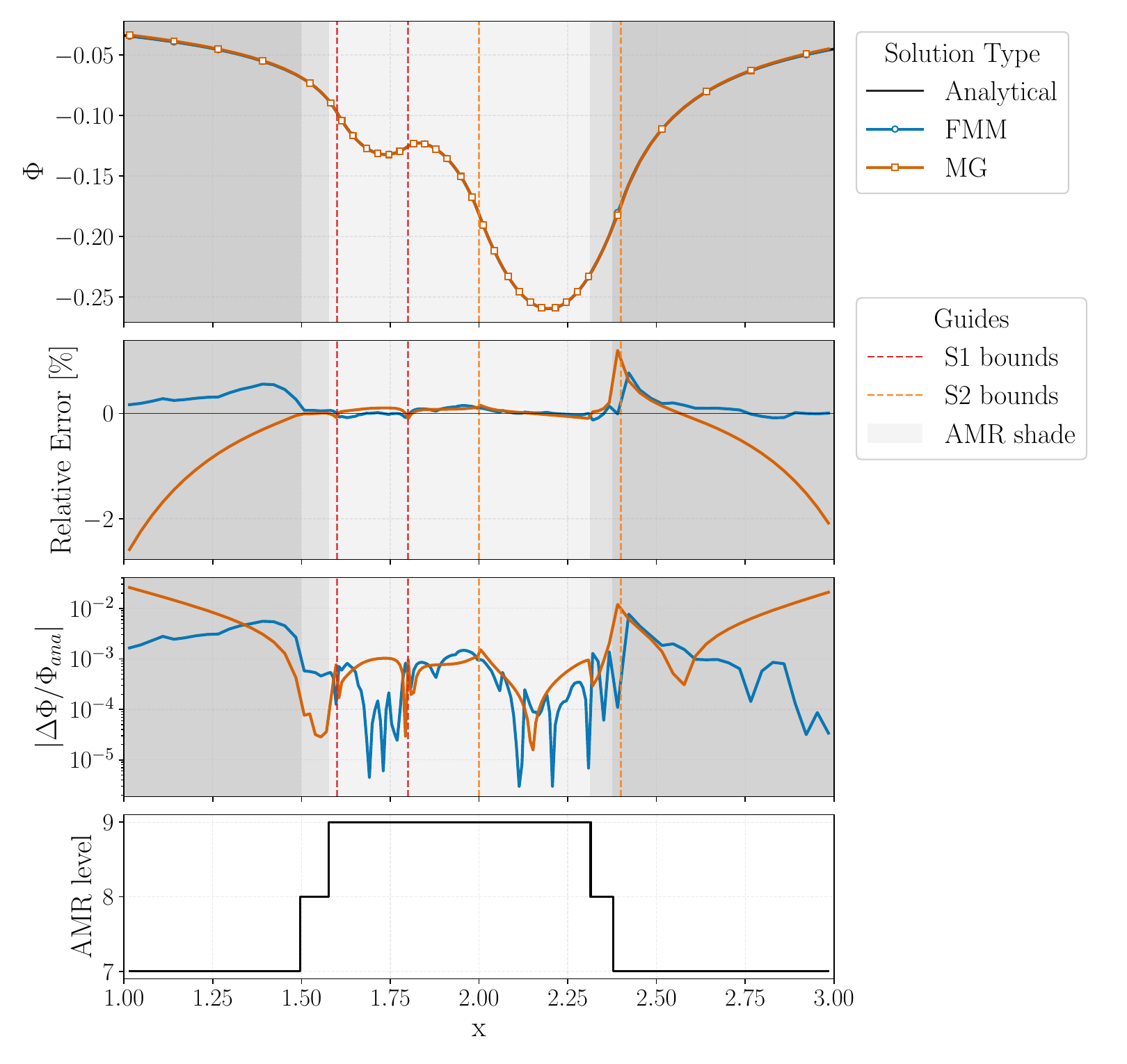}
    \caption{A double uniform spheres test case shown in Paper I, now with geometrically refined AMR grid. We display the profile of gravitational potentials obtained using FMM (\textit{blue}) and MG (\textit{orange}) in comparison with the analytical solution (\textit{black}). The values are calculated at the cell-centers closest to the axis piercing through the spheres' centers. \textit{Dotted vertical lines} guide the interiors of the two spheres, and the \textit{gray shades} indicate the AMR level, darker being coarser.}
    \label{fig:spheres}
\end{figure*}

\subsection{Overview of the algorithm: The Life Cycle of FMM trees} \label{sec:lifecycle}

In this section, we provide pseudo-codes for the full FMM algorithm with AMR and ATS. RAMSES realizes ATS through sub-cycling, meaning that finer grid levels are advanced using smaller time steps. This behavior can be specified independently for each level. When sub-cycling is active, the finer level is updated by performing two leap-frog iterations, each using a time step equal to half of the coarser level’s time step. Thus, when solving for the system at a given level, only the finer refinement levels are synchronized and are referred to as \textit{active levels}. It is only the active levels that could be further refined or de-refined. The life cycle of the tree should correspond to the sub-cycling scheme, and thus trees should be rebuilt to reflect the new AMR structure, as well as the new mass distribution. Subsequently, we rebuild the merged tree for the active levels. Since our trees are sparsely built on top of the refinement regions, we rebuild the FMM trees for active levels at each fine time step, unlike other codes that approximates by drifting the multipoles according to the center of mass velocity and reconstructing the full tree only after a fixed cumulative portion of particles receive a force update \citep[e.g.,][]{GADGET-4,PKDGRAV3,SWIFT}.

Below, we provide an overview of the FMM algorithm for level $\ell$. First, let $\mathcal{T}_j$ denote the FMM tree at level $j$,
$\mathbf{M}_j$ the corresponding multipole coefficients, and $\mathbf{L}_\ell$ the local expansion at level $\ell$. For $\ell<\ell_{\rm max}$, define the merged tree
$\tilde{\mathcal{T}}_{\ell+1}
= \mathrm{Merge} \left(
\{\mathcal{T}_j\}_{j=\ell+1}^{\ell_{\rm max}}
\right)$. The variable \texttt{icount} holds the information if it is in the first or the second half of sub-cycling. The FMM trees are rebuilt for the active levels if we are solving for the base grid, or if we are at the second half of sub-cycling. 

\vspace{0.3em}
\noindent \textbf{Algorithm 1:} \textsc{FMM Solve}($\ell$, \texttt{icount})
\begin{algorithmic}[1]
\If{$(\ell=\ell_{\rm min})\lor(\texttt{icount}>1)$}

    \State \Call{RebuildTree}{$\ell$}

\EndIf

\If{$\ell<\ell_{\rm max}$}

    \State \Call{RebuildMergedTree}{$\ell$}

    \State
    $\tilde{\mathcal{T}}_{\ell+1}
    \gets
    \mathrm{Merge}
    (\{\mathcal{T}_j\}_{j=\ell+1}^{\ell_{\rm max}})$

\EndIf

\State $\mathbf{L}_\ell\gets0$

\State \Call{MultipoleToLocal}{$\ell$}

\State \Call{PotentialEvaluation}{$\ell$}

\State Apply $\mathbf{L}_\ell$

\end{algorithmic}

At each rebuild, the stale active trees are replaced with new FMM trees with refreshed multipole information. As illustrated in Figure~\ref{fig:tree}, \textsc{BuildTree} efficiently constructs the complete hierarchical FMM grid solely in regions where AMR cells are actually present. The \textsc{Upward} routine aggregates the multipoles, which are later shifted to the cell's centers using \textsc{CenterShift}. Merged tree is destroyed and rebuilt at every FMM solve call through \textsc{rebuild merged}, which is similar to the standard {rebuild} routine without \textsc{Upward} and \textsc{CenterShift}. Consecutively, the multipoles from $\tilde{\mathbf{M}}_{j=\ell+1}^{\ell_{\rm max}}$ are aggregated to $\tilde{\mathbf{M}}_j$. The additional merging operation costs $\mathcal{O}(N_{\rm active})$, making the algorithm still linear and efficient. 

\vspace{0.3em}
\noindent \textbf{Algorithm 2:} \textsc{RebuildTree}($\ell$)
\begin{algorithmic}[1]
\State discard $\mathcal{T}_j$
for $j=\ell,\dots,\ell_{\rm max}$

\For{$j=\ell$ to $\ell_{\rm max}$}

    \State
    $\mathcal{T}_j
    \gets$
    \Call{BuildTree}{$j$}

    \State
    $\mathbf{M}_j
    \gets$
    \Call{Upward}{$\mathcal{T}_j$}

\EndFor

\For{$j=\ell$ to $\ell_{\rm max}$}
    \State
    $\mathbf{M}_j
    \gets$
    \Call{CenterShift}
    {$\mathcal{T}_j,\mathbf{M}_j$}
\EndFor
\end{algorithmic}

Next, we describe how the downward pass is performed. Here, let us use the variable $f$ to reference the level of the FMM grid within a designated tree (i.e., $\mathcal{T}_{\ell';f}$). Let $\mathbf{L}_{\ell;f}$ and $\mathbf{M}_{\ell;f}$ denote, respectively, the local expansions and multipole moments associated with the FMM grid at level $f$ of the tree $\mathcal{T}_{\ell}$. If $\ell=f$, the multipoles correspond to the mass of the AMR cells as the FMM grid that store the full multipoles starts one level coarser than the AMR grid. Finally, let $\mathcal{K}_{j\to\ell}$ denote the operation that uses the multipole or mass information to calculate local expansions or the potential. If a full multipole input is given, it will act as either M2P or M2L, and P2P if given the mass input. As described below, the local expansions are shifted by the L2L operation, \textsc{ShiftLocal}, to account for the far field. For evaluation of either local expansion or potential, we iterate through all possible FMM trees. However, given the refinement rules of RAMSES (graded octree), only up to $\ell-1$ for coarser cells should be considered. For finer levels $\ell'>\ell$, the merged tree's multipoles are used if they exist (i.e., $\ell<\ell_{\rm max}$).

\vspace{0.3em}
\noindent \textbf{Algorithm 3:} \textsc{MultipoleToLocal}($\ell$)
\begin{algorithmic}[1]
\For{$f=\ell_{\rm bound}+1$ to $\ell-1$}
    \State $\mathbf{L}_{\ell;f} \gets \mathrm{ShiftLocal}( \mathbf{L}_{\ell;f-1})$

    \For{$j=\max(\ell_{\rm min},f-1)$ to $\min(\ell,\ell_{\rm max})$}

        \State
        $\mathbf{L}_{\ell;f} \mathrel{+}=
        \mathcal{K}_{j\to\ell}\mathbf{M}_{j;f-1}$

    \EndFor

    \If{$\ell<\ell_{\rm max}$}

        \State
        $\mathbf{L}_{\ell;f} \mathrel{+}=
        \mathcal{K}_{\tilde{\ell+1}\to\ell}
        \tilde{\mathbf{M}}_{\ell+1;f-1}$

    \EndIf

\EndFor

\end{algorithmic}

\vspace{0.3em}
\noindent \textbf{Algorithm 4:} \textsc{PotentialEvaluation}($\ell$)
\begin{algorithmic}[1]
\State \texttt{Far field}
\State $\Phi_\ell \gets \mathrm{ShiftLocal}( \mathbf{L}_{\ell;\ell-1})$
\Statex
\State \texttt{Intermediate field}

\If{$\ell<\ell_{\rm max}$}

    \State
    $\Phi_\ell \mathrel{+}=
    \mathcal{K}_{\tilde{\ell+1}\to\ell}
    \tilde{\mathbf{M}}_{\ell+1;\ell-1}$

\Else

    \State
        $\Phi_\ell \mathrel{+}=
        \mathcal{K}_{j\to\ell_{\rm max}}\mathbf{M}_{\ell_{\rm max};\ell_{\rm max}-1}$

\EndIf
\Statex
\State \texttt{Near and Nearest fields}

\For{$j=\max(\ell_{\rm min},\ell-1)$ to $\ell$}

    \State
    $\Phi_\ell \mathrel{+}=
    \mathcal{K}_{j\to\ell} M_j$

\EndFor

\If{$\ell<\ell_{\rm max}$}

    \State
    $\Phi_\ell \mathrel{+}=
    \mathcal{K}_{\tilde{\ell+1}\to\ell}
    \tilde{\mathbf{M}}_{\ell+1}$

\Else

    $\Phi_\ell \mathrel{+}=
        \mathcal{K}_{j\to\ell_{\rm max}}\mathbf{M}_{\ell_{\rm max};\ell_{\rm max}-1}$

\EndIf

\end{algorithmic}

\begin{figure*}
	\includegraphics[width=\textwidth]{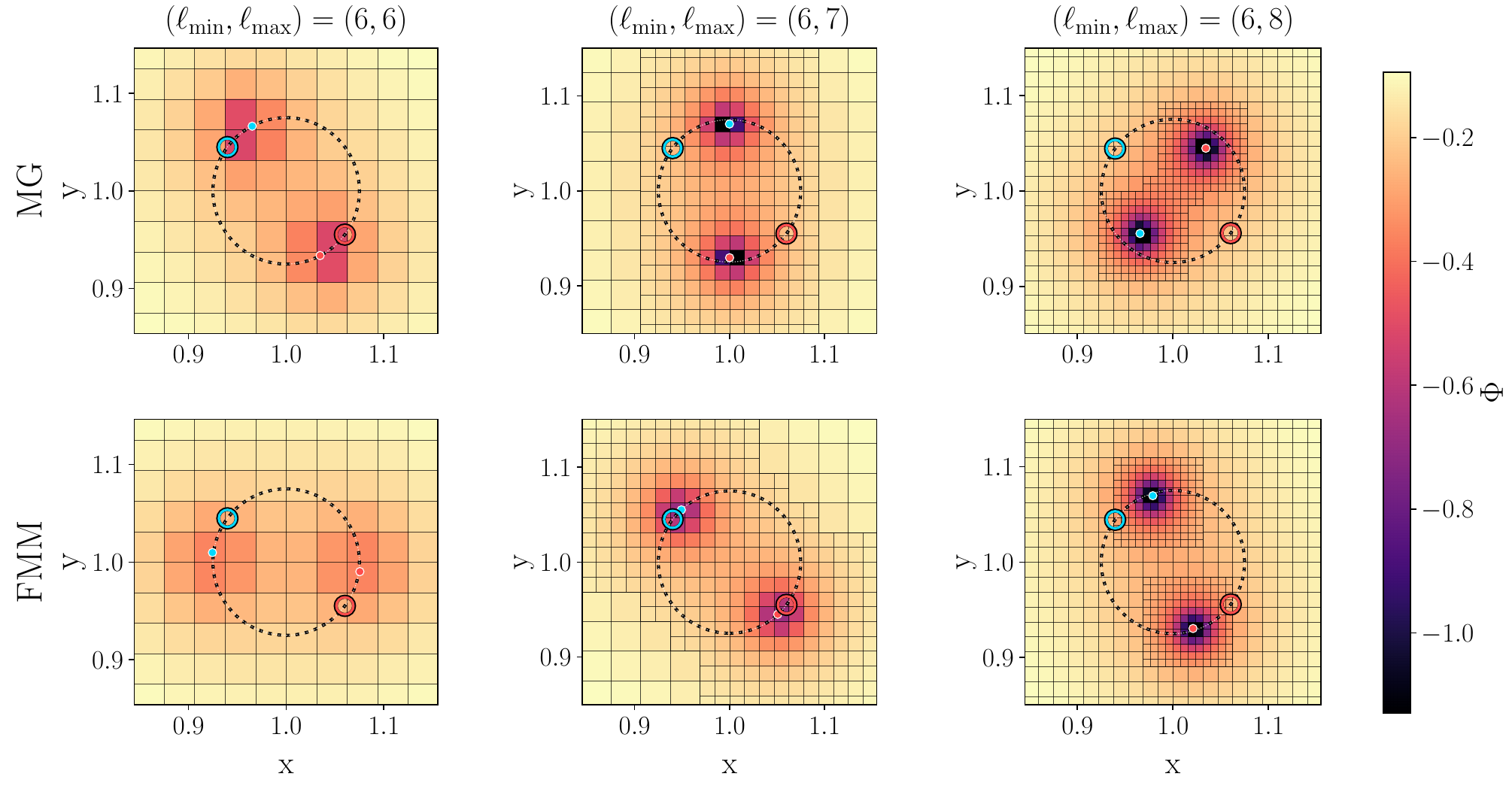}
    \caption{A test problem consisting of two isolated, equal-mass point charges orbiting one another. The charges, with $GM=0.01$, are separated by $0.15$ and placed inside a box of length 1 to reduce boundary-induced biases. The system is evolved up to $t=8$, which corresponds to roughly three orbits. In each panel, the analytical solution is plotted with \textit{hollow circles}, while the numerical results are indicated by \textit{filled circles}. The \textit{large dashed circles} serve as guides to the true circular orbit. The \textit{top row} presents results obtained with MG and the \textit{bottom row} with FMM. We vary the AMR configuration so that $\ell_{\rm max}$ increases from 6 to 8, while keeping $\ell_{\rm min}=6$ fixed. The charges are always located at the finest refinement level, and the timestep is held constant to eliminate any influence from ATS, thereby preserving the symplectic nature of the leap-frog integrator. Although both solvers show good agreement with the analytical solution at unigrid (despite the phase shift), MG fails to retain the original circular orbit and starts to spiral in for deeper AMR grids.}
    \label{fig:binary}
\end{figure*}

\begin{figure*}
\includegraphics[width=\textwidth]{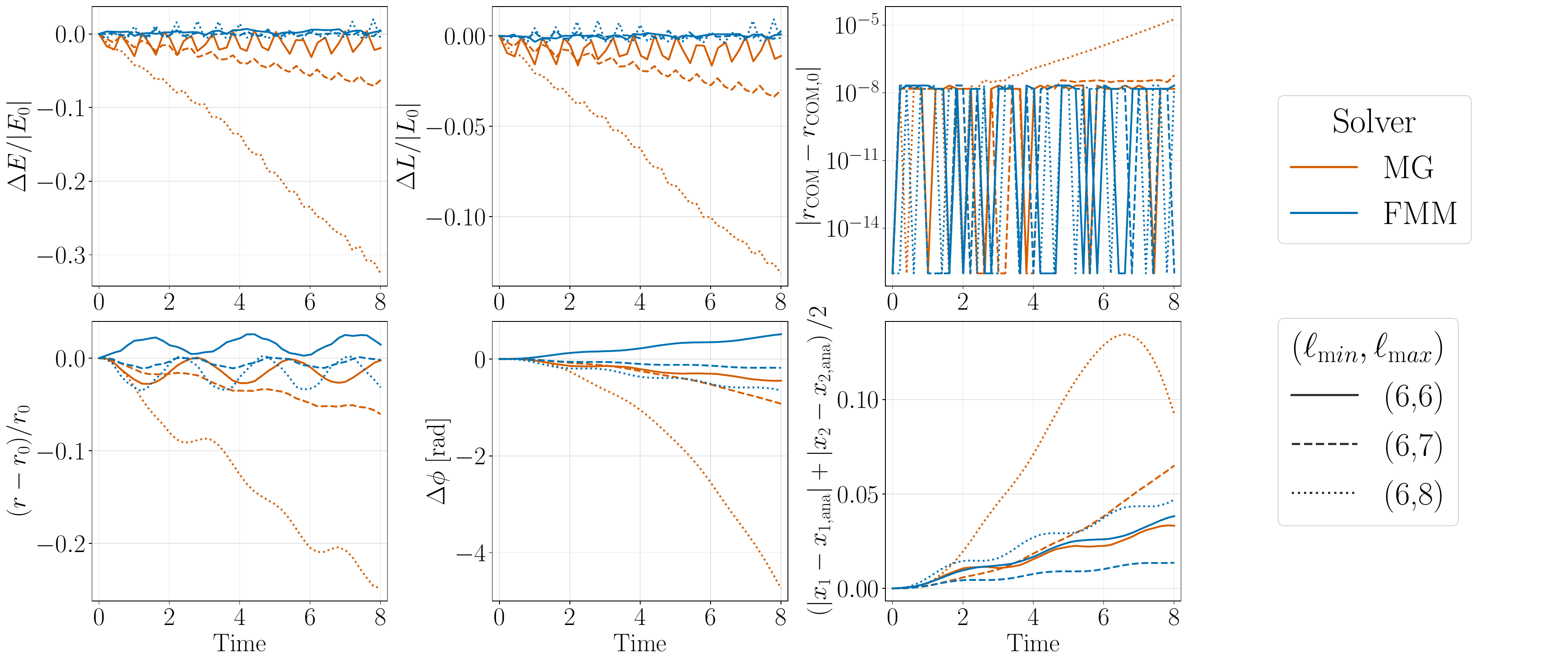}
    \caption{Diagnostics of conservation properties for the binary orbit test case in Fig.~\ref{fig:binary}. From top left to bottom right, the panels compare the relative difference of total Energy ($\Delta E/E_0$), angular momentum ($\Delta L/L_0$), center of mass drift ($|r_{\rm COM}| - r_{\rm COM,0}|$), interparticle distance ($(r-r_0)/r_0$), phase drift ($\Delta \phi$), and the mean absolute deviation of position from the analytical solutions ($(|x_1 - x_{1,\rm ana}|+|x_2 - x_{2,\rm ana}|)/2$). Results for FMM are shown in \textit{blue}, and MG in \textit{orange}, with \textit{solid lines} for $(\ell_{\rm min}, \ell_{\rm max})=(6,6)$, \textit{dashed lines} for $(6,7)$ and \textit{dotted lines} for $(6,8)$.}
    \label{fig:binary_diagnostics}
\end{figure*}

\begin{figure*}
	\includegraphics[width=0.9\textwidth]{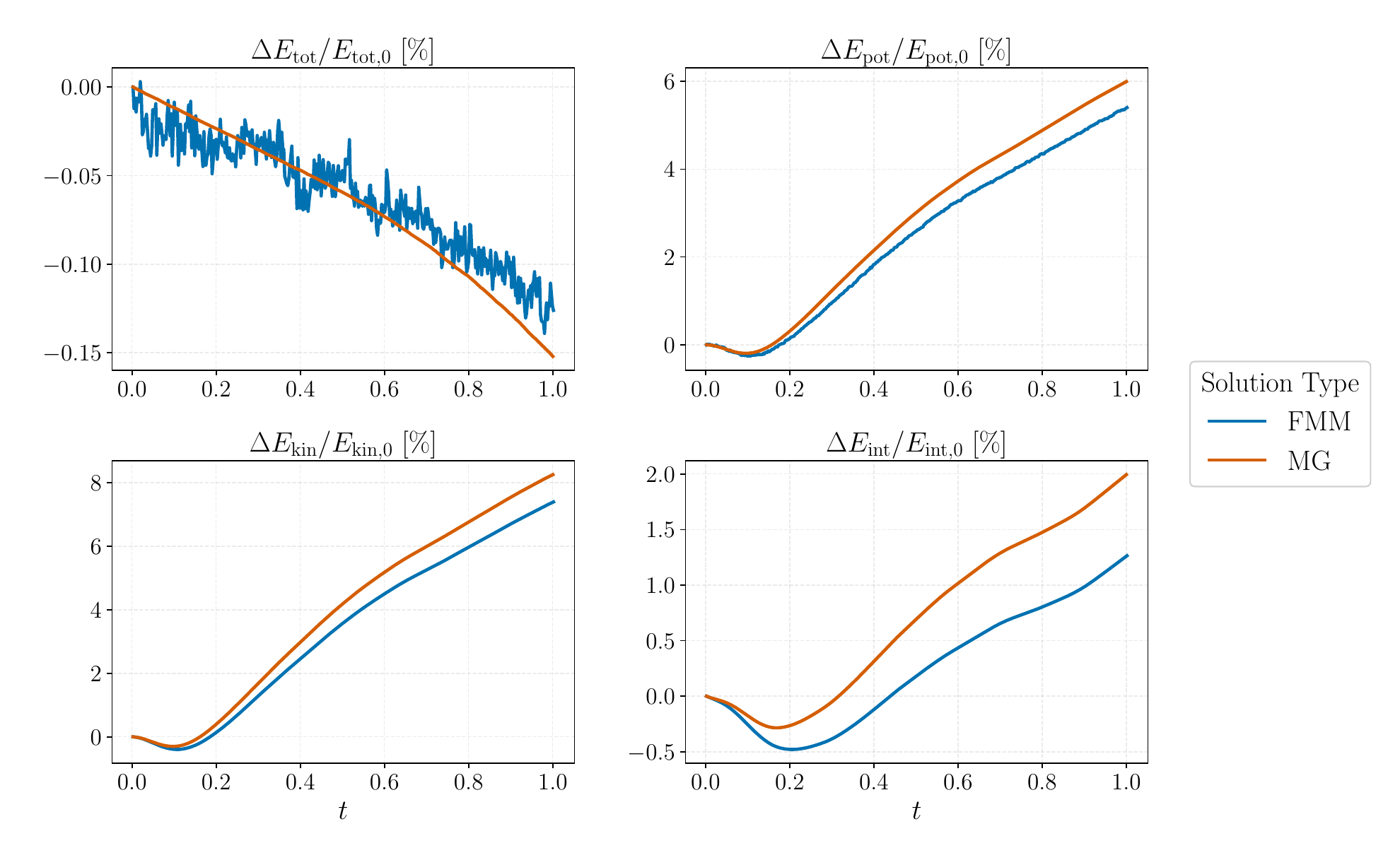}
    \caption{Mass and energy conservation diagnostics together with relative changes in each energy budget for an isolated and hydrostatic NFW halo ($v_{\rm 200}=150$~km/s and $c=10$) with gas  (mass fraction of $f_{\rm gas}=15\%$) for FMM (\textit{blue}) and MG (\textit{orange}) solved with AMR and ATS. Both solvers demonstrate stable solutions at similar levels for a long period ($t_{\rm unit}\approx 0.98$~Gyr.) The small fluctuations from the relative difference in total energy partly arise from changes in the potential energy, which are caused by differences in how the field is sampled as refined regions move. In contrast, the relative difference of kinetic and internal energies increase smoothly, indicating that the dynamics remain stable against the potential perturbations introduced by the box-like error patterns of FMM.}
    \label{fig:nfw_diagnostics}
\end{figure*}

\section{Test Results}
\label{sec:test}
In this section, we present several test cases to verify the FMM algorithm in both AMR and ATS, compared to MG. Unless otherwise indicated, we use the standard parameter values for MG: a tolerance of $\epsilon = 10^{-6}$, which usually results in about 6 V-cycles.

Figure~\ref{fig:nfw} compares an isolated NFW halo's gravitational potential map on a fixed four-level AMR grid. The agreement between FMM and MG is excellent, with a mean relative difference of $0.51\%$ and a maximum difference of only $1.55\%$. No noticeable discontinuities are observed at AMR interfaces. The usual box-like error structures associated with the intrinsic discontinuity of local expansions across FMM levels are present, apart from the largest source of error at the central peak due to the mass deposition operator onto the grid.

Figure~\ref{fig:spheres} reexamines the double sphere test case from Paper I, where we test the gravitational potential field from two uniform density spheres with radii 0.1 and 0.2. The centers of the sphere are located at $(0.7, 1.0, 1.0)$ and $(1.2, 1.0, 1.0)$, inside a cubic box of length 2.0. We extend this problem to the AMR case by geometrically refining the region in between, as shown by the gray shades in the figure. The typical bias from the inexact Dirichlet boundary conditions is seen near the box boundaries for MG. Again, the discontinuity of the solutions primarily arises from the density discontinuity rather than at the refinement level boundaries, showing robustness of both solvers, and the two exhibit similar levels of peakiness throughout.

To evaluate how well the two gravity solvers conserve momentum, we consider a simple setup in which two equal-mass point charges orbit each other on a circular trajectory. The two charges, each with $GM=0.01$, are placed at a separation of $r=0.15$, with the center of mass located in the center of the box. The computational domain has size 1, large enough to eliminate boundary-induced artifacts. To examine the impact of AMR, we perform three simulations with the base grid fixed at $\ell_{\rm min}=6$ while varying the maximum refinement level as $\ell_{\rm max}=6,7,8$. The particles are always forced to reside on the finest available refinement level. Lastly, we use a fixed time step in order to maintain the symplectic nature of the leap-frog integrator, and no sub-cycling is performed.

Figure~\ref{fig:binary} shows the final configuration of the two charges at $t=8$. The filled circles mark the numerical solution, and the hollow circles the analytic solution. For the unigrid case, both MG and FMM show good match despite some phase shift. However, as we move on to deeper AMR levels, the solution from MG breaks down as two particles spiral inwards, whereas the FMM's solution is in good agreement with the analytical solution off to some phase shift. The deviation for MG is especially pronounced in the case $(\ell_{\rm min},\ell_{\rm max})=(6,8)$, where the two particles are in disconnected refined regions, and interpolation across the coarse-fine interface strongly affects the solution.

Figure~\ref{fig:binary_diagnostics} presents several diagnostics for the binary orbit test case: the relative change in total energy ($\Delta E/E_0$), angular momentum ($\Delta L/L_0$), center-of-mass drift ($|r_{\rm COM} - r_{{\rm COM},0}|$), interparticle separation ($(r-r_0)/r_0$), phase drift ($\Delta \phi$), and the mean absolute deviation of the particle positions from the analytical solutions ($(|x_1 - x_{1,\rm ana}|+|x_2 - x_{2,\rm ana}|)/2$). For all diagnostics, FMM consistently produces more accurate solutions than MG. Furthermore, for FMM the deviation in interparticle separation and from the analytical solution does not necessarily grow with increasing $\ell_{\rm max}$. In contrast, MG clearly exhibits increasing deviations as the grid is refined, and the solutions deteriorate quickly at higher refinement levels. Therefore, this test case shows that FMM has improved conservation properties compared to MG even with the boxy error patterns intrinsic to FMM.

\begin{figure*}
	\includegraphics[width=\textwidth]{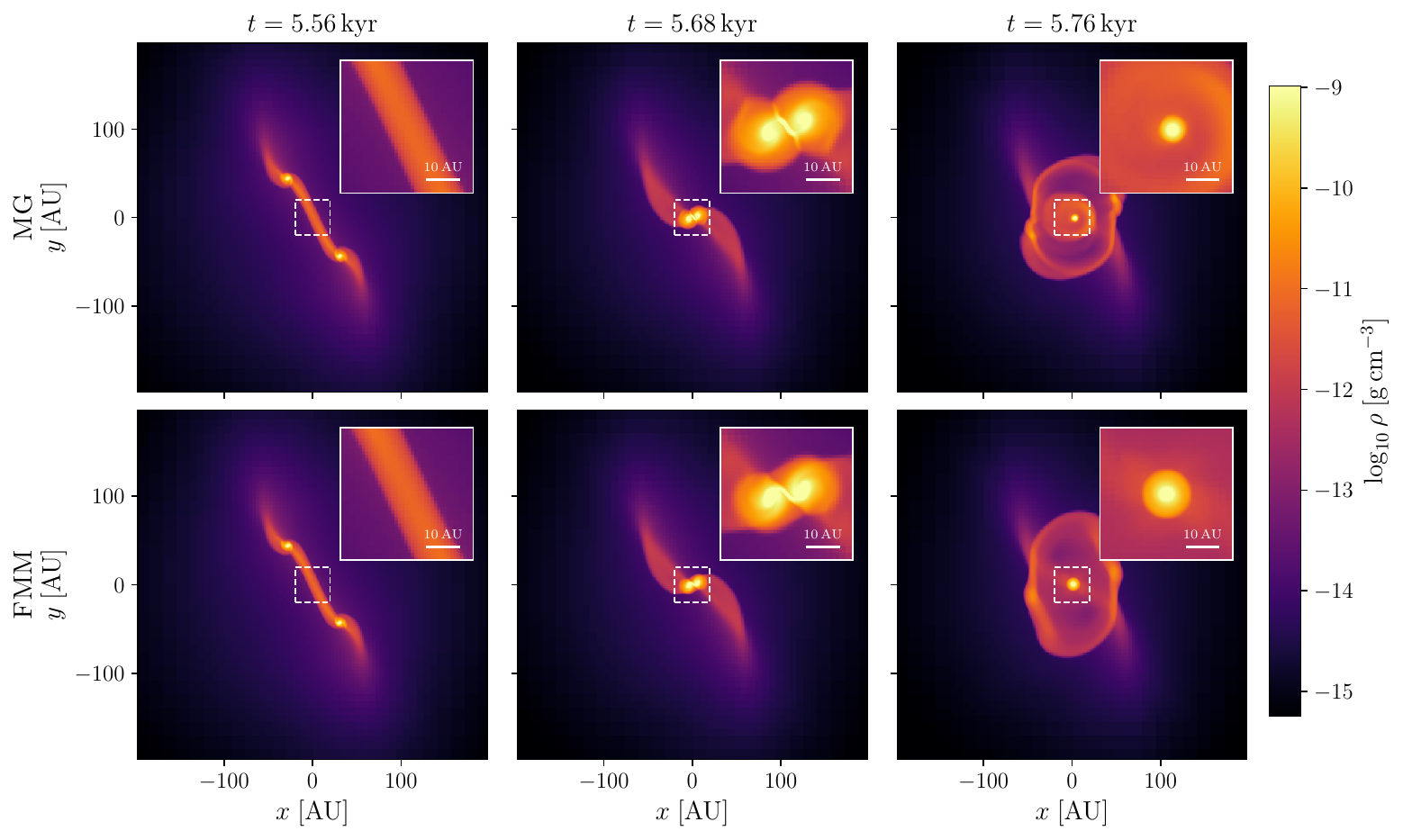}
    \caption{Comparison between MG (\textit{top row}) and FMM (\textit{bottom row}) solutions for the \citet{Boss_Bodenheimer_1979} test, which describes the protostellar collapse from a rotating cloud. The collapse begins from a spherical cloud of uniform density with a mass of 1$\,M_{\rm \odot}$ and a radius of about $1000\,\mathrm{AU}$, with a $10\%$ perturbation of the $m=2$ mode. Initial rotation is set so that the rotational energy is 1\% of the gravitational potential energy. The gas evolves according to the barotropic equation of state described in \citet{Machida_2006} and \citet{Marchand_2016}, starting from an initial temperature of $10\,\mathrm{K}$. The figures and their inset panels each show the gas column density at the center, zoomed in by 10 and 100 times, respectively. From \textit{left} to \textit{right}, we show periods when two spiral overdensities form, collide, and merge to form the first Larson core \citep{Larson_1969}. The global morphologies are in good agreement, with differences arising only in the precise structures of the collision fronts and the first Larson core within its surrounding environment.}
    \label{fig:collapse}
\end{figure*}

The failing mode of MG can be attributed again to the inexact Dirichlet boundary conditions, but across the coarse-fine level interface in this case. Although subtler than the bias seen at the box boundaries, the use of Dirichlet boundary conditions does not generally enforce the continuity of the gravitational potential and the conservation of force fluxes across coarse-fine interfaces, leading to spurious accelerations\footnote{One should note that this error from coarse-fine force fluxes mismatch is different from the violation of the curl-free condition of gravity, which can arise when gravity is coupled to hydrodynamics, as discussed for example by \citet{Mullen2021}.}.

The first effect---discontinuity from the interpolated potential field across the coarse-fine interface---can have a great impact especially if the body is too close to the interface, and necessitates a sufficient volume of refinement region around the bodies\footnote{Thus in the experiments shown in Figures~\ref{fig:binary} and \ref{fig:binary_diagnostics}, we perform mesh smoothing with $\texttt{nexpand}=2$ at every refinement level to prevent the particles from grazing the coarse-fine interface too close. Without mesh expansion, the results for MG deteriorates even more while FMM stays rather robust to the degree of mesh expansion. Section \ref{sec:nexpand} shows the identical experiments performed with $\texttt{nexpand}=1$, which demonstrates the importance of sufficient mesh smoothing for MG.}. Secondly, in order to correct for the force flux mismatch, one needs to conserve both the Neumann and Dirichlet boundary conditions across the coarse-fine interface, which is often referred to as elliptic matching \citep{MartinCartwright1996}. This is somewhat similar in spirit to the flux correction across AMR cells in hydrodynamics, as in \citet{BergerColella}. However, this treatment also causes non-trivial problems when extended to the ATS scheme, as elliptic matching should be performed as a synchronized operation, while in ATS each level is advanced with its own sub-cycles. For example, \textsc{Gamer-I} performs flux correction across coarse-fine interfaces but disables ATS when self-gravity is enabled \citep{GamerI}. For \textsc{Gamer-II}, on the other hand, ATS is enabled, but at the cost of bypassing exact elliptic matching and mitigating the resulting boundary errors by adding buffer zones \citep{GamerII}. Likewise, \textsc{Athena++} enforces normal-gradient matching across level boundaries through a conservative discretization of the Laplacian, but adopts uniform time stepping \citep{2023ApJS..266....7T}.

Many existing FMM implementations enforce exact momentum conservation using a dual-tree walk formalism, which ensures that interactions between source and sink bodies are treated mutually and therefore generate equal and opposite forces \citep{Dehnen2000_symmetry,Dehnen2002_theory,Bernard2026}. This is realized by checking pairwise the opening angle, or the multipole acceptance criteria, between cell-cell or cell-body pairs. Our design adheres rather to the rigid geometric AMR-like opening angle criteria, as discussed in Section~\ref{sec:nearest} and Figure~\ref{fig:field}, which does not observe strict symmetry. However, we ensure the mutuality between coarse-fine interfaces via the nearest field decomposition where the forces are most influential. In the binary orbit test case, the sink and source cells are not connected through the nearest-field treatment, but instead separated by the intermediate or far-field of the level it resides in. The orbit should therefore remain sensitive to the intrinsic box-like error patterns of FMM. Nevertheless, the system remains stable, indicating that these residual errors are sufficiently small and do not induce significant secular drift in this configuration.

Finally, similar to the synchronization issue encountered in elliptic matching for MG, the coarse inactive multipoles are not drifted during each sub-cycle and are therefore subject to a temporal synchronization error. For example, \citet{SWIFT} stores the center-of-mass velocities of cells and drifts the multipoles, achieving first-order accuracy in time. We therefore also test the binary problem with ATS for both MG and FMM. Although the FMM solution shows some deviation compared to the fixed-time-step case, partly due to the loss of the symplectic property of the time integrator, it still maintains a stable and well-matched orbit. In contrast, MG starts to spiral inward in a similar manner. In principle, the algorithm could be redesigned to enforce stricter conservation of total momentum employing the dual-tree walk or multipole drifting. However, following the rigid geometry given by the mesh refinement structure enables the precomputation of translation kernels, which reduces arithmetic operations and makes the algorithm well suited for GPU acceleration. Multipole drifting can be incorporated with additional memory and computational cost and will be explored in future work. Moreover RAMSES is mainly used for applications including cosmological $N$-body simulations, where Poissonian noise is unavoidable, and self-gravitating astrophysical fluids, for which the current level of accuracy and conservation appears sufficient for the intended applications, unlike $N$-body codes for stellar dynamics that need more careful treatment and higher polynomial orders \citep[e.g.,][]{Dehnen2014_stellar_dynamics}.

To further examine the validity of the adaptive FMM scheme, we performed a controlled test of the hydrostatic stability of an isolated NFW halo in AMR and ATS. The halo has properties of $v_{200}=150~{\rm km~s^{-1}}$ and concentration $c=10$, and contains gas with mass fraction $f_{\rm gas}=0.15$. Figure~\ref{fig:nfw_diagnostics} shows the mass conservation error, the energy conservation diagnostic based on the relative variations of the potential ($E_{\rm pot}$), kinetic ($E_{\rm kin}$), internal ($E_{\rm int}$), and total energies ($E_{\rm tot}=E_{\rm kin}+E_{\rm pot}+E_{\rm int}$) over approximately $0.98~{\rm Gyr}$. Both solvers preserve hydrostatic equilibrium to a similar degree. The small fluctuations in the energy conservation diagnostics may have two origins. First, they may reflect genuine small changes in the dynamics as the gas and particles are slightly kicked from the discontinuities. Second, the AMR hierarchy samples different FMM regions as the AMR structure evolves, and this sampling can introduce small discontinuous changes in the measured energy. Indeed, the fluctuations are visible from the potential energy rather than from kinetic or internal energies, which hints that the effect does not result in the change in dynamics but rather sampling of the discontinuous field maps. Moreover, these fluctuations are much smaller than the secular relative deviation from the initial state, remain comparable in amplitude to those seen with MG. This experiment shows that the small kicks occurring at FMM level boundaries, caused by discontinuities in the local expansions, do not substantially degrade the accuracy of the hydrostatic solution.

\begin{figure*}
	\includegraphics[width=\textwidth]{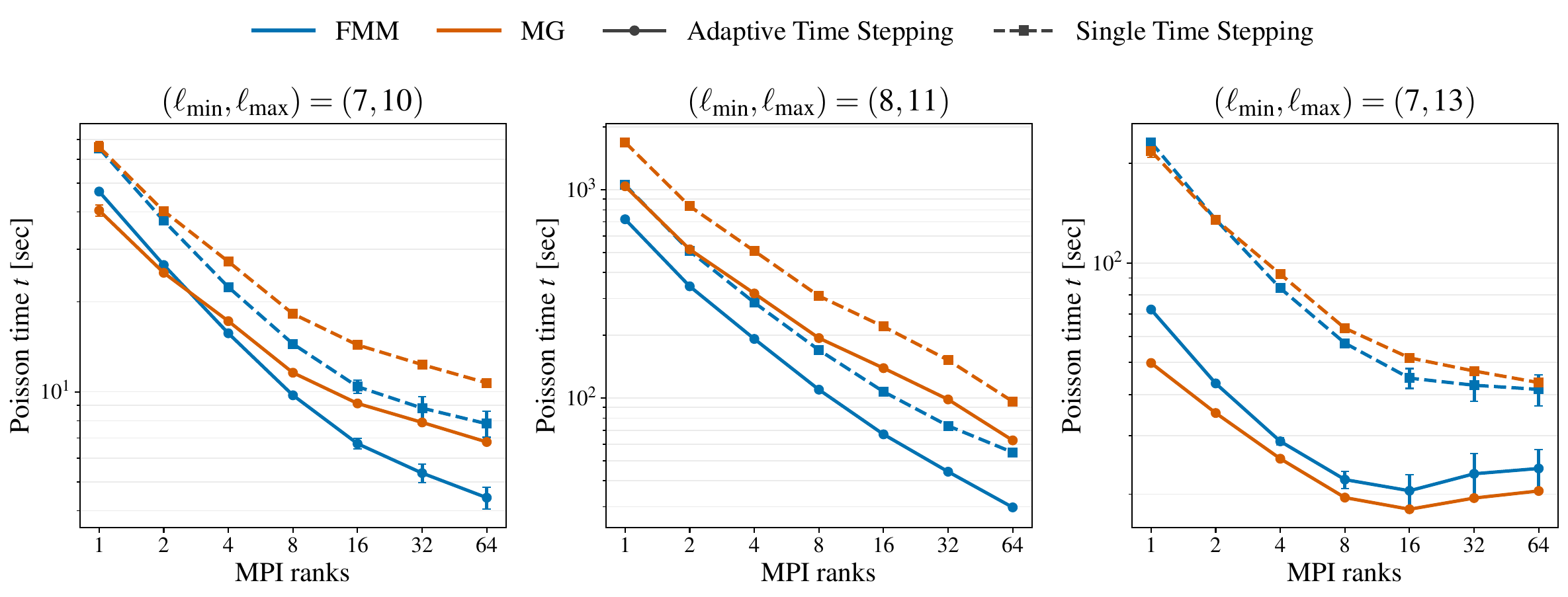}
    \caption{Strong scaling results comparing FMM (\textit{blue}) and MG (\textit{dark orange}) run with adaptive time stepping (\textit{solid lines}) and with single time stepping (\textit{dashed lines}). We evolve an NFW halo using grid sizes of $(\ell_{\mathrm{min}},\ell_{\mathrm{max}})=(7,10)$ (\textit{left panel}), $(8,11)$ (\textit{middle panel}), and $(7,13)$ (\textit{right panel}) up to a fixed final simulation time. For each grid configuration, we solve the problem with increasing number of MPI ranks. The experiments are repeated ten times per configuration, and the error bars are shown.}
    \label{fig:scaling}
\end{figure*}

Finally, we also test dynamical problems that include both hydrodynamics and cooling in AMR and ATS, namely a protostellar collapse. We first test the \citet{Boss_Bodenheimer_1979} model of protostellar collapse without magnetic fields, starting from a uniform density spherical cloud of mass $1\,M_\odot$ and radius of approximately 1000 AU. An initial $m=2$ density perturbation with an amplitude of 10\% is imposed, and the cloud undergoes slow rotation such that its rotational energy corresponds to 1\% of its gravitational potential energy. The gas is initially cold, with $T=10\,\mathrm{K}$, and evolves according to a piecewise barotropic equation of state following \citet{Machida_2006} and \citet{Marchand_2016}: an initially isothermal collapse, adiabatic heating during first Larson core formation \citep{Larson_1969}, softening due to $\mathrm{H}_2$ dissociation, and final adiabatic stiffening after dissociation as the second Larson core is formed. The simulation box size is set as approximately 4000 AU (four times the initial radius) with a very deep hierarchy of AMR grids of twelve levels, or an effective finest resolution of $0.24\,\mathrm{AU}$ to resolve up to the first Larson core. Finally, we impose a Truelove condition of the Jeans refinement ($\Delta x_{\rm min}\leq 8 \lambda_J$) \citep{Truelove1997}.

In Figure~\ref{fig:collapse}, we compare the solutions we obtain from MG and FMM at the same times, depicting the formation of two overdensities, which eventually collide along the connecting ridges and merge to form the first Larson core. The overall global morphologies match very well, showing robustness of the two different gravity solvers. However, we also observe minor differences in the morphology of the collision front, which result in a slightly different core size and alter the morphology of the surrounding ambient medium. We find that despite the MG's disadvantages in terms of strict momentum conservation, the overall evolution of astrophysical systems of interest does not seem to be impacted. 

\section{Performance Scaling}
\label{sec:scaling}

In this section, we compare the scaling behaviors of MG and FMM with AMR and ATS. Figure \ref{fig:scaling} demonstrates the strong scaling behaviors with ATS and with single time stepping for each of the solvers in a variety of grid configurations. We evolve a dark-matter-only NFW halo with initial conditions generated from \textsc{Dice} \citep{dice} over a fixed final simulation time. To assess how AMR grid structure affects scaling, we compare the default setup, $(\ell_{\mathrm{min}},\ell_{\mathrm{max}})=(7,10)$, against $(8,11)$, where oct counts are relatively balanced across levels, and $(7,13)$, where the hierarchy is deeper and the finest levels are sparsely populated. Each experiment is conducted ten times, and the resulting timing errors are shown with error bars.

Overall, independent of gravity solver, ATS with sub-cycling provides a speed-up of roughly $\times 1.5$ for the cases $(7,10)$ and $(8,11)$, where the grid structure is relatively well balanced across refinement levels. For the much deeper $(7,13)$ hierarchy, the gain increases to factors of $\sim 2$--$3$. This is because, without sub-cycling, the global timestep is set by the sparsely populated finest level, forcing all coarser levels to advance at the time step of the finest level even though most of the computational work resides on coarser levels. Comparing the two solvers, MG is slightly faster on a single MPI rank for the $(7,10)$ and $(7,13)$ cases, but FMM shows better overall parallel scaling and rapidly becomes faster as the number of MPI ranks increases. The single-rank advantage of MG is consistent with Paper I for the unigrid results, where the FMM algorithm requires roughly 30 times more arithmetic operations than MG, but nevertheless reaches a comparable wall-clock time because both solvers operate largely in the memory-bound regime. The $(8,11)$ case provides the clearest AMR example of this effect. Since it contains many more octs overall, distributed more evenly across refinement levels, the fixed overheads of FMM tree construction, traversal, and near-field bookkeeping are better amortized. As a result, FMM already outperforms MG even on a single MPI rank.

The performance gain of FMM over MG increases as the number of MPI ranks increases, and this advantage becomes more pronounced when each AMR level contains many octs per rank, as shown by the case $(8,11)$. The $(7,13)$ case demonstrates an opposite behavior, where the finer levels are under-populated, especially with only a few grids at the finest level. The early saturation of scaling at 16 MPI ranks for $(7,13)$ indeed demonstrates that the well-populated grid structure is crucial for speed-up. However, even in such extreme conditions, FMM is fairly robust and performs better than MG.

\section{Conclusions}
\label{sec:conclusion}
In this work, we extend the previous unigrid algorithm for FMM into an adaptive algorithm compatible with the AMR and ATS framework of RAMSES for isolated boundary conditions. Our implementation of FMM in AMR is unique in two ways. First, unlike most implementations of FMM directly acting on particles, our algorithm works on the adaptive mesh where the mass from particles and gas are deposited onto (Adaptive Particle Mesh). Second, instead of employing opening angle or multipole acceptance criteria, as in many tree codes, we adhere to the native AMR structure. This design is advantageous because the mesh structure of RAMSES can closely follow the local physics through fine-grained, cell-based refinement, while the refinement criteria can be flexibly adjusted according to the target problem. By following this refined AMR structure directly, our method naturally inherits these advantages. An additional contribution of this work is a direct, apples-to-apples comparison between two distinct Poisson solvers, FMM and MG, at comparable accuracy and for the same AMR structure.

A key feature of the adaptive algorithm is the use of multiple FMM trees, each assigned to a specific refinement level and responsible for storing the multipole contribution of the corresponding level. By doing this, we can retain the information from coarser inactive levels, while advancing the finer active levels. Trees with stale multipole information are refreshed whenever the FMM algorithm restarts from the coarsest refinement level, or during the second step of the two-step sub-cycling scheme. Next, following the refinement rules of the graded octree in RAMSES, we construct a general interaction list applicable to arbitrary AMR geometries. In particular, to enforce force symmetry across coarse–fine boundaries in the near field, we introduce the nearest field, where coarse and fine cells interact mutually. Furthermore, to expedite redundant neighborhood searches from different FMM trees, we introduce a merged tree, where the multipole information from active levels are merged.

Next, we perform several test cases and compare FMM against MG on identical meshes. The two solvers show excellent agreement for the NFW halo test, with sub-percent differences in the gravitational potential profile. The double-sphere test further confirms the characteristic box-like error patterns of FMM, caused by discontinuities in the local expansions, as well as the bias near the box boundaries in MG, which arises from the propagation of inexact Dirichlet boundary conditions. Moreover, we find similar peaky qualities of errors across the solutions, showing that the two solvers are indeed behaving very similarly. 

We highlight one distinct advantage of FMM over MG by testing two isolated, equal-mass point charges orbiting one another. While FMM conserves energy and (angular) momentum of the system very well, MG fails to retain stability and the particles spirals inward, especially when the particles reside in highly refined regions and are disconnected by coarser cells. We attribute this to the spurious accelerations at coarse-fine interfaces from the inexact Dirichlet boundary conditions. In order to mitigate this, \textit{(i)} sufficient volume of refinement region around the particles to prevent particles from grazing the interface too close via mesh smoothing and \textit{(ii)} elliptic matching that conserves both Dirichlet and Neumann boundary conditions are necessary \citep{MartinCartwright1996}. However, elliptic matching in the presence of ATS is non-trivial, as the levels are out of synchronization during sub-cycling. 

FMM shows stable numerical results in diverse mesh configurations, despite the corrections needed above. Although the rigid interaction geometry of our FMM implementation does not guarantee strict momentum conservation in the same sense as a dual-tree prescription, we enforce force symmetry in the near-field interactions. Moreover, our tests suggest that the discontinuities between local expansions do not significantly compromise the conservation properties of the method. This is further demonstrated by the evolution of a hydrostatic NFW halo, where FMM and MG conserve energy to comparable accuracy even in the presence of sub-cycling. A dynamic test of a protostellar collapse from a rotating cloud \citep{Boss_Bodenheimer_1979} also exhibits similar global evolution of the system, and shows the full capacity of the two distinct solvers for realistic self-gravitating astrophysical fluids.

Finally, we measure the strong scaling performance of the two solvers. First, we find that for both solvers, sub-cycling provides a speed-up of roughly a factor of 1.5. Furthermore, although performance on a single MPI rank slightly favors MG, especially when the AMR hierarchy is deep and sparse, FMM scales better as the number of MPI ranks increases. This trend is robust across the different AMR structures tested, suggesting that the scaling advantage of FMM is not specific to a particular refinement geometry. This is consistent with the finding of Paper I, where FMM performs about 30 times more operations than MG in the unigrid case, but benefits from higher arithmetic intensity and more effective memory reuse. The performance gain of FMM over MG is enhanced when the number of octs per level is balanced and well populated, but the better scaling persists even for deeper and sparser AMR hierarchies.

The present implementation is restricted to isolated boundary conditions. Extending the method to periodic domains will be crucial for cosmological simulations. 
An interesting possibility is a hybrid PM-FMM approach, where the long-range force is computed with PM and the short-range force with a truncated FMM kernel \citep{Wang2021}.

Our FMM implementation in RAMSES, which is natively compatible with AMR and ATS, will be useful for a broad range of astrophysical fluid simulations in which self-gravity plays an important role. Given the robust momentum conservation behavior demonstrated in the binary orbit test, the method should be especially well suited to systems with dynamically evolving and geometrically complex AMR structures. 

Finally, the algorithm was designed with future optimization for GPU acceleration in mind: Cartesian expansions based on cell centers allow fixed interaction geometries and precomputed translation kernels, while the relatively low expansion order keeps the memory footprint modest. In addition, the comparatively high arithmetic intensity of FMM suggests that it may be well suited to GPU architectures, which are particularly effective for compute-intensive algorithms. These properties make the method a promising foundation for future high-performance AMR gravity solvers in preparation for the exascale era.

\section*{Acknowledgements}

J.-Y.L thanks Nickolay Y. Gnedin and David Velasco-Romero for insightful discussion. This material is based upon work supported by the National Science Foundation (NSF) and the
U.S.-Israel Binational Science Foundation (BSF) under Award Number 2406558 and Award Title ``The Origin of the Excess of Bright Galaxies at Cosmic Dawn''. The authors are also pleased to acknowledge that the work reported in this paper was performed substantially using Princeton University’s Research Computing resources, specifically the Stellar cluster.

\section*{Data Availability}

The implementation of FMM in the prototype code {\ttfamily mini-ramses} is available from the authors upon reasonable request.



\bibliographystyle{mnras}
\bibliography{example} 

@ARTICLE{FMM_Unigrid_RAMSES,
       author = {{Lee}, Jun-Young and {Teyssier}, Romain},
        title = "{A Scalable Fast Multipole Method Poisson Solver for the RAMSES code: I. Unigrid Algorithm}",
      journal = {\mnras},
         year = 2026,
        month = jun,
          doi = {10.1093/mnras/stag1241},
archivePrefix = {arXiv},
       eprint = {2606.13793},
 primaryClass = {astro-ph.IM},
       adsurl = {https://ui.adsabs.harvard.edu/abs/2026MNRAS.tmp.1151L}
}

@ARTICLE{Gingold_Monaghan_1977,
       author = {{Gingold}, R.~A. and {Monaghan}, J.~J.},
        title = "{Smoothed particle hydrodynamics: theory and application to non-spherical stars.}",
      journal = {\mnras},
         year = 1977,
        month = nov,
       volume = {181},
        pages = {375-389},
          doi = {10.1093/mnras/181.3.375},
       adsurl = {https://ui.adsabs.harvard.edu/abs/1977MNRAS.181..375G}
}

@ARTICLE{Berger_Oliger_1984,
       author = {{Berger}, Marsha J. and {Oliger}, Joseph},
        title = "{Adaptive Mesh Refinement for Hyperbolic Partial Differential Equations}",
      journal = {Journal of Computational Physics},
         year = 1984,
        month = mar,
       volume = {53},
       number = {3},
        pages = {484-512},
          doi = {10.1016/0021-9991(84)90073-1},
       adsurl = {https://ui.adsabs.harvard.edu/abs/1984JCoPh..53..484B}
}

@inproceedings{Brandt1977MultilevelAS,
  title={Multi-level adaptive solutions to boundary-value problems math comptr},
  author={Achi Brandt},
  year={1977},
  url={https://api.semanticscholar.org/CorpusID:5919303}
}

@article{FEDORENKO19621092,
	author = {R.P. Fedorenko},
	doi = {https://doi.org/10.1016/0041-5553(62)90031-9},
	issn = {0041-5553},
	journal = {USSR Computational Mathematics and Mathematical Physics},
	number = {4},
	pages = {1092-1096},
	title = {A relaxation method for solving elliptic difference equations},
	url = {https://www.sciencedirect.com/science/article/pii/0041555362900319},
	volume = {1},
	year = {1962}}

@article{BAKHVALOV1966101,
	author = {N.S. Bakhvalov},
	doi = {https://doi.org/10.1016/0041-5553(66)90118-2},
	issn = {0041-5553},
	journal = {USSR Computational Mathematics and Mathematical Physics},
	number = {5},
	pages = {101-135},
	title = {On the convergence of a relaxation method with natural constraints on the elliptic operator},
	url = {https://www.sciencedirect.com/science/article/pii/0041555366901182},
	volume = {6},
	year = {1966}}

@ARTICLE{RAMSES_2002,
       author = {{Teyssier}, R.},
        title = "{Cosmological hydrodynamics with adaptive mesh refinement. A new high resolution code called RAMSES}",
      journal = {\aap},
         year = 2002,
        month = apr,
       volume = {385},
        pages = {337-364},
          doi = {10.1051/0004-6361:20011817},
archivePrefix = {arXiv},
       eprint = {astro-ph/0111367},
 primaryClass = {astro-ph},
       adsurl = {https://ui.adsabs.harvard.edu/abs/2002A&A...385..337T}
}

@ARTICLE{GUILLET_2011,
       author = {{Guillet}, Thomas and {Teyssier}, Romain},
        title = "{A simple multigrid scheme for solving the Poisson equation with arbitrary domain boundaries}",
      journal = {Journal of Computational Physics},
         year = 2011,
        month = jun,
       volume = {230},
       number = {12},
        pages = {4756-4771},
          doi = {10.1016/j.jcp.2011.02.044},
archivePrefix = {arXiv},
       eprint = {1104.1703},
 primaryClass = {physics.comp-ph},
       adsurl = {https://ui.adsabs.harvard.edu/abs/2011JCoPh.230.4756G}
}

@ARTICLE{PKDGRAV3,
       author = {{Potter}, Douglas and {Stadel}, Joachim and {Teyssier}, Romain},
        title = "{PKDGRAV3: beyond trillion particle cosmological simulations for the next era of galaxy surveys}",
      journal = {Computational Astrophysics and Cosmology},
         year = 2017,
        month = may,
       volume = {4},
       number = {1},
          eid = {2},
        pages = {2},
          doi = {10.1186/s40668-017-0021-1},
archivePrefix = {arXiv},
       eprint = {1609.08621},
 primaryClass = {astro-ph.IM},
       adsurl = {https://ui.adsabs.harvard.edu/abs/2017ComAC...4....2P}
}

@article{GADGET-4,
    author = {Springel, Volker and Pakmor, Rüdiger and Zier, Oliver and Reinecke, Martin},
    title = "{Simulating cosmic structure formation with the gadget-4 code}",
    journal = {Monthly Notices of the Royal Astronomical Society},
    volume = {506},
    number = {2},
    pages = {2871-2949},
    year = {2021},
    month = {07},
    issn = {0035-8711},
    doi = {10.1093/mnras/stab1855},
    url = {https://doi.org/10.1093/mnras/stab1855},
    eprint = {https://academic.oup.com/mnras/article-pdf/506/2/2871/39271725/stab1855.pdf},
}

@ARTICLE{Dehnen2002_theory,
       author = {{Dehnen}, Walter},
        title = "{A Hierarchical <E10>O</E10>(N) Force Calculation Algorithm}",
      journal = {Journal of Computational Physics},
         year = 2002,
        month = jun,
       volume = {179},
       number = {1},
        pages = {27-42},
          doi = {10.1006/jcph.2002.7026},
archivePrefix = {arXiv},
       eprint = {astro-ph/0202512},
 primaryClass = {astro-ph},
       adsurl = {https://ui.adsabs.harvard.edu/abs/2002JCoPh.179...27D}
}

@ARTICLE{Dehnen2000_symmetry,
       author = {{Dehnen}, Walter},
        title = "{A Very Fast and Momentum-conserving Tree Code}",
      journal = {\apjl},
         year = 2000,
        month = jun,
       volume = {536},
       number = {1},
        pages = {L39-L42},
          doi = {10.1086/312724},
archivePrefix = {arXiv},
       eprint = {astro-ph/0003209},
 primaryClass = {astro-ph},
       adsurl = {https://ui.adsabs.harvard.edu/abs/2000ApJ...536L..39D}
}

@ARTICLE{Greengard_Rokhlin_1987,
       author = {{Greengard}, L. and {Rokhlin}, V.},
        title = "{A Fast Algorithm for Particle Simulations}",
      journal = {Journal of Computational Physics},
         year = 1987,
        month = dec,
       volume = {73},
       number = {2},
        pages = {325-348},
          doi = {10.1016/0021-9991(87)90140-9},
       adsurl = {https://ui.adsabs.harvard.edu/abs/1987JCoPh..73..325G}
}

@ARTICLE{1997AcNum...6..229G,
       author = {{Greengard}, Leslie and {Rokhlin}, Vladimir},
        title = "{A new version of the Fast Multipole Method for the Laplace equation in three dimensions}",
      journal = {Acta Numerica},
         year = 1997,
        month = jan,
       volume = {6},
        pages = {229-269},
          doi = {10.1017/S0962492900002725},
       adsurl = {https://ui.adsabs.harvard.edu/abs/1997AcNum...6..229G}
}

@ARTICLE{Cheng_Greengard_Rokhlin_1999,
       author = {{Cheng}, H. and {Greengard}, L. and {Rokhlin}, V.},
        title = "{A Fast Adaptive Multipole Algorithm in Three Dimensions}",
      journal = {Journal of Computational Physics},
         year = 1999,
        month = nov,
       volume = {155},
       number = {2},
        pages = {468-498},
          doi = {10.1006/jcph.1999.6355},
       adsurl = {https://ui.adsabs.harvard.edu/abs/1999JCoPh.155..468C}
}

@article{Hrycak_Rokhlin1998,
	author = {Hrycak, Tomasz and Rokhlin, Vladimir},
	doi = {10.1137/S106482759630989X},
	eprint = {https://doi.org/10.1137/S106482759630989X},
	journal = {SIAM Journal on Scientific Computing},
	number = {6},
	pages = {1804-1826},
	title = {An Improved Fast Multipole Algorithm for Potential Fields},
	url = {https://doi.org/10.1137/S106482759630989X},
	volume = {19},
	year = {1998}}

@ARTICLE{Greengard_Rokhlin_laplace_3d,
       author = {{Greengard}, Leslie and {Rokhlin}, Vladimir},
        title = "{A new version of the Fast Multipole Method for the Laplace equation in three dimensions}",
      journal = {Acta Numerica},
         year = 1997,
        month = jan,
       volume = {6},
        pages = {229-269},
          doi = {10.1017/S0962492900002725},
       adsurl = {https://ui.adsabs.harvard.edu/abs/1997AcNum...6..229G}
}

@ARTICLE{Dehnen2014_stellar_dynamics,
       author = {{Dehnen}, Walter},
        title = "{A fast multipole method for stellar dynamics}",
      journal = {Computational Astrophysics and Cosmology},
         year = 2014,
        month = sep,
       volume = {1},
          eid = {1},
        pages = {1},
          doi = {10.1186/s40668-014-0001-7},
archivePrefix = {arXiv},
       eprint = {1405.2255},
 primaryClass = {astro-ph.IM},
       adsurl = {https://ui.adsabs.harvard.edu/abs/2014ComAC...1....1D}
}

@ARTICLE{Barnes_Hut_1986,
       author = {{Barnes}, Josh and {Hut}, Piet},
        title = "{A hierarchical O(N log N) force-calculation algorithm}",
      journal = {\nat},
         year = 1986,
        month = dec,
       volume = {324},
       number = {6096},
        pages = {446-449},
          doi = {10.1038/324446a0},
       adsurl = {https://ui.adsabs.harvard.edu/abs/1986Natur.324..446B}
}

@BOOK{Hockney_Eastwood_1988,
       author = {{Hockney}, R.~W. and {Eastwood}, J.~W.},
        title = "{Computer simulation using particles}",
         year = 1988,
       adsurl = {https://ui.adsabs.harvard.edu/abs/1988csup.book.....H}
}

@ARTICLE{2023ApJS..266....7T,
       author = {{Tomida}, Kengo and {Stone}, James M.},
        title = "{The Athena++ Adaptive Mesh Refinement Framework: Multigrid Solvers for Self-gravity}",
      journal = {\apjs},
         year = 2023,
        month = may,
       volume = {266},
       number = {1},
          eid = {7},
        pages = {7},
          doi = {10.3847/1538-4365/acc2c0},
archivePrefix = {arXiv},
       eprint = {2302.13903},
 primaryClass = {astro-ph.IM},
       adsurl = {https://ui.adsabs.harvard.edu/abs/2023ApJS..266....7T}
}

@ARTICLE{Ethridge&Greengrad2001,
       author = {{Ethridge}, Frank and {Greengard}, Leslie},
        title = "{A New Fast-Multipole Accelerated Poisson Solver in Two Dimensions}",
      journal = {SIAM Journal on Scientific Computing},
         year = 2001,
        month = jan,
       volume = {23},
       number = {3},
        pages = {741-760},
          doi = {10.1137/S1064827500369967},
       adsurl = {https://ui.adsabs.harvard.edu/abs/2001SJSC...23..741E}
}

@software{dice,
       author = {{Perret}, Valentin},
        title = "{DICE: Disk Initial Conditions Environment}",
 howpublished = {Astrophysics Source Code Library, record ascl:1607.002},
         year = 2016,
        month = jul,
          eid = {ascl:1607.002},
archivePrefix = {ascl},
       eprint = {1607.002},
       adsurl = {https://ui.adsabs.harvard.edu/abs/2016ascl.soft07002P}
}

@ARTICLE{SWIFT,
       author = {{Schaller}, Matthieu and {Borrow}, Josh and {Draper}, Peter W. and {Ivkovic}, Mladen and {McAlpine}, Stuart and {Vandenbroucke}, Bert and {Bah{\'e}}, Yannick and {Chaikin}, Evgenii and {Chalk}, Aidan B.~G. and {Chan}, Tsang Keung and {Correa}, Camila and {van Daalen}, Marcel and {Elbers}, Willem and {Gonnet}, Pedro and {Hausammann}, Lo{\"\i}c and {Helly}, John and {Hu{\v{s}}ko}, Filip and {Kegerreis}, Jacob A. and {Nobels}, Folkert S.~J. and {Ploeckinger}, Sylvia and {Revaz}, Yves and {Roper}, William J. and {Ruiz-Bonilla}, Sergio and {Sandnes}, Thomas D. and {Uyttenhove}, Yolan and {Willis}, James S. and {Xiang}, Zhen},
        title = "{SWIFT: A modern highly-parallel gravity and smoothed particle hydrodynamics solver for astrophysical and cosmological applications}",
      journal = {\mnras},
         year = 2024,
        month = may,
       volume = {530},
       number = {2},
        pages = {2378-2419},
          doi = {10.1093/mnras/stae922},
archivePrefix = {arXiv},
       eprint = {2305.13380},
 primaryClass = {astro-ph.IM},
       adsurl = {https://ui.adsabs.harvard.edu/abs/2024MNRAS.530.2378S}
}

@ARTICLE{ART,
       author = {{Kravtsov}, Andrey V. and {Klypin}, Anatoly A. and {Khokhlov}, Alexei M.},
        title = "{Adaptive Refinement Tree: A New High-Resolution N-Body Code for Cosmological Simulations}",
      journal = {\apjs},
         year = 1997,
        month = jul,
       volume = {111},
       number = {1},
        pages = {73-94},
          doi = {10.1086/313015},
archivePrefix = {arXiv},
       eprint = {astro-ph/9701195},
 primaryClass = {astro-ph},
       adsurl = {https://ui.adsabs.harvard.edu/abs/1997ApJS..111...73K}
}

@ARTICLE{Castro,
       author = {{Almgren}, A.~S. and {Beckner}, V.~E. and {Bell},
                      J.~B. and {Day}, M.~S. and {Howell}, L.~H. and
                      {Joggerst}, C.~C. and {Lijewski}, M.~J. and
                      {Nonaka}, A. and {Singer}, M. and {Zingale}, M.},
        title = "{CASTRO: A New Compressible Astrophysical
                      Solver. I. Hydrodynamics and Self-gravity}",
      journal = {\apj},
    archivePrefix = "arXiv",
       eprint = {1005.0114},
     primaryClass = "astro-ph.IM",
         year = 2010,
        month = jun,
       volume = 715,
        pages = {1221-1238},
          doi = {10.1088/0004-637X/715/2/1221},
       adsurl = {http://adsabs.harvard.edu/abs/2010ApJ...715.1221A}
    }

@ARTICLE{Enzo,
   author = {{Bryan}, G.~L. and {Norman}, M.~L. and {O'Shea}, B.~W. and {Abel}, T. and 
   {Wise}, J.~H. and {Turk}, M.~J. and {Reynolds}, D.~R. and {Collins}, D.~C. and 
   {Wang}, P. and {Skillman}, S.~W. and {Smith}, B. and {Harkness}, R.~P. and 
   {Bordner}, J. and {Kim}, J.-h. and {Kuhlen}, M. and {Xu}, H. and 
   {Goldbaum}, N. and {Hummels}, C. and {Kritsuk}, A.~G. and {Tasker}, E. and 
   {Skory}, S. and {Simpson}, C.~M. and {Hahn}, O. and {Oishi}, J.~S. and 
   {So}, G.~C. and {Zhao}, F. and {Cen}, R. and {Li}, Y. and {The Enzo Collaboration}
   },
    title = "{ENZO: An Adaptive Mesh Refinement Code for Astrophysics}",
  journal = {\apjs},
archivePrefix = "arXiv",
   eprint = {1307.2265},
 primaryClass = "astro-ph.IM",
     year = 2014,
    month = apr,
   volume = 211,
      eid = {19},
    pages = {19},
      doi = {10.1088/0067-0049/211/2/19},
   adsurl = {http://adsabs.harvard.edu/abs/2014ApJS..211...19B}
}

@ARTICLE{GamerI,
       author = {{Schive}, Hsi-Yu and {Tsai}, Yu-Chih and {Chiueh}, Tzihong},
        title = "{GAMER: A Graphic Processing Unit Accelerated Adaptive-Mesh-Refinement Code for Astrophysics}",
      journal = {\apjs},
         year = 2010,
        month = feb,
       volume = {186},
       number = {2},
        pages = {457-484},
          doi = {10.1088/0067-0049/186/2/457},
archivePrefix = {arXiv},
       eprint = {0907.3390},
 primaryClass = {astro-ph.IM},
       adsurl = {https://ui.adsabs.harvard.edu/abs/2010ApJS..186..457S}
}

@ARTICLE{GamerII,
       author = {{Schive}, Hsi-Yu and {ZuHone}, John A. and {Goldbaum}, Nathan J. and {Turk}, Matthew J. and {Gaspari}, Massimo and {Cheng}, Chin-Yu},
        title = "{GAMER-2: a GPU-accelerated adaptive mesh refinement code - accuracy, performance, and scalability}",
      journal = {\mnras},
         year = 2018,
        month = dec,
       volume = {481},
       number = {4},
        pages = {4815-4840},
          doi = {10.1093/mnras/sty2586},
archivePrefix = {arXiv},
       eprint = {1712.07070},
 primaryClass = {astro-ph.IM},
       adsurl = {https://ui.adsabs.harvard.edu/abs/2018MNRAS.481.4815S}
}

@ARTICLE{OctoTiger,
       author = {{Marcello}, Dominic C. and {Shiber}, Sagiv and {De Marco}, Orsola and {Frank}, Juhan and {Clayton}, Geoffrey C. and {Motl}, Patrick M. and {Diehl}, Patrick and {Kaiser}, Hartmut},
        title = "{OCTO-TIGER: a new, 3D hydrodynamic code for stellar mergers that uses HPX parallelization}",
      journal = {\mnras},
         year = 2021,
        month = jul,
       volume = {504},
       number = {4},
        pages = {5345-5382},
          doi = {10.1093/mnras/stab937},
archivePrefix = {arXiv},
       eprint = {2101.08226},
 primaryClass = {astro-ph.IM},
       adsurl = {https://ui.adsabs.harvard.edu/abs/2021MNRAS.504.5345M}
}

@techreport{MartinCartwright1996,
  author      = {Martin, D. F. and Cartwright, K. L.},
  title       = {Solving Poisson's Equation Using Adaptive Mesh Refinement},
  institution = {EECS Department, University of California, Berkeley},
  number      = {UCB/ERL M96/66},
  year        = {1996},
  month       = oct,
  url         = {https://www2.eecs.berkeley.edu/Pubs/TechRpts/1996/3105.html}
}

@article{BergerColella,
	author = {M.J. Berger and P. Colella},
	doi = {https://doi.org/10.1016/0021-9991(89)90035-1},
	issn = {0021-9991},
	journal = {Journal of Computational Physics},
	number = {1},
	pages = {64-84},
	title = {Local adaptive mesh refinement for shock hydrodynamics},
	url = {https://www.sciencedirect.com/science/article/pii/0021999189900351},
	volume = {82},
	year = {1989}}

@ARTICLE{Mullen2021,
       author = {{Mullen}, P.~D. and {Hanawa}, Tomoyuki and {Gammie}, C.~F.},
        title = "{An Extension of the Athena++ Framework for Fully Conservative Self-gravitating Hydrodynamics}",
      journal = {\apjs},
         year = 2021,
        month = feb,
       volume = {252},
       number = {2},
          eid = {30},
        pages = {30},
          doi = {10.3847/1538-4365/abcfbd},
archivePrefix = {arXiv},
       eprint = {2012.01340},
 primaryClass = {astro-ph.IM},
       adsurl = {https://ui.adsabs.harvard.edu/abs/2021ApJS..252...30M}
}

@ARTICLE{Truelove1997,
       author = {{Truelove}, J. Kelly and {Klein}, Richard I. and {McKee}, Christopher F. and {Holliman}, II, John H. and {Howell}, Louis H. and {Greenough}, Jeffrey A.},
        title = "{The Jeans Condition: A New Constraint on Spatial Resolution in Simulations of Isothermal Self-gravitational Hydrodynamics}",
      journal = {\apjl},
         year = 1997,
        month = nov,
       volume = {489},
       number = {2},
        pages = {L179-L183},
          doi = {10.1086/310975},
       adsurl = {https://ui.adsabs.harvard.edu/abs/1997ApJ...489L.179T}
}

@ARTICLE{Wang2021,
       author = {{Wang}, Qiao},
        title = "{A hybrid Fast Multipole Method for cosmological N-body simulations}",
      journal = {Research in Astronomy and Astrophysics},
         year = 2021,
        month = jan,
       volume = {21},
       number = {1},
          eid = {003},
        pages = {003},
          doi = {10.1088/1674-4527/21/1/3},
archivePrefix = {arXiv},
       eprint = {2006.14952},
 primaryClass = {physics.comp-ph},
       adsurl = {https://ui.adsabs.harvard.edu/abs/2021RAA....21....3W}
}

@ARTICLE{Bernard2026,
       author = {{Bernard}, Yann and {David-Cl{\'e}ris}, Timoth{\'e}e and {Price}, Daniel J. and {Lau}, Mike Y.~M.},
        title = "{Momentum-conserving self-gravity in the phantom smoothed particle hydrodynamics code. Parallel dual tree traversal for the symmetric fast multipole method}",
      journal = {arXiv e-prints},
         year = 2026,
        month = feb,
          eid = {arXiv:2602.05804},
        pages = {arXiv:2602.05804},
          doi = {10.48550/arXiv.2602.05804},
archivePrefix = {arXiv},
       eprint = {2602.05804},
 primaryClass = {astro-ph.IM},
       adsurl = {https://ui.adsabs.harvard.edu/abs/2026arXiv260205804B}
}

@ARTICLE{Boss_Bodenheimer_1979,
       author = {{Boss}, A.~P. and {Bodenheimer}, P.},
        title = "{Fragmentation in a rotating protostar: a comparison of two three-dimensional computer codes.}",
      journal = {\apj},
         year = 1979,
        month = nov,
       volume = {234},
        pages = {289-295},
          doi = {10.1086/157497},
       adsurl = {https://ui.adsabs.harvard.edu/abs/1979ApJ...234..289B}
}

@ARTICLE{Machida_2006,
       author = {{Machida}, Masahiro N. and {Inutsuka}, Shu-ichiro and {Matsumoto}, Tomoaki},
        title = "{Second Core Formation and High-Speed Jets: Resistive Magnetohydrodynamic Nested Grid Simulations}",
      journal = {\apjl},
         year = 2006,
        month = aug,
       volume = {647},
       number = {2},
        pages = {L151-L154},
          doi = {10.1086/507179},
archivePrefix = {arXiv},
       eprint = {astro-ph/0603456},
 primaryClass = {astro-ph},
       adsurl = {https://ui.adsabs.harvard.edu/abs/2006ApJ...647L.151M}
}

@ARTICLE{Marchand_2016,
       author = {{Marchand}, P. and {Masson}, J. and {Chabrier}, G. and {Hennebelle}, P. and {Commer{\c{c}}on}, B. and {Vaytet}, N.},
        title = "{Chemical solver to compute molecule and grain abundances and non-ideal MHD resistivities in prestellar core-collapse calculations}",
      journal = {\aap},
         year = 2016,
        month = jul,
       volume = {592},
          eid = {A18},
        pages = {A18},
          doi = {10.1051/0004-6361/201526780},
archivePrefix = {arXiv},
       eprint = {1604.05613},
 primaryClass = {astro-ph.GA},
       adsurl = {https://ui.adsabs.harvard.edu/abs/2016A&A...592A..18M}
}

@ARTICLE{Larson_1969,
       author = {{Larson}, Richard B.},
        title = "{Numerical calculations of the dynamics of collapsing proto-star}",
      journal = {\mnras},
         year = 1969,
        month = jan,
       volume = {145},
        pages = {271},
          doi = {10.1093/mnras/145.3.271},
       adsurl = {https://ui.adsabs.harvard.edu/abs/1969MNRAS.145..271L}
}



\appendix
\section{Binary Orbit Test without Mesh Expansion}\label{sec:nexpand}
\begin{figure*}
	\includegraphics[width=\textwidth]{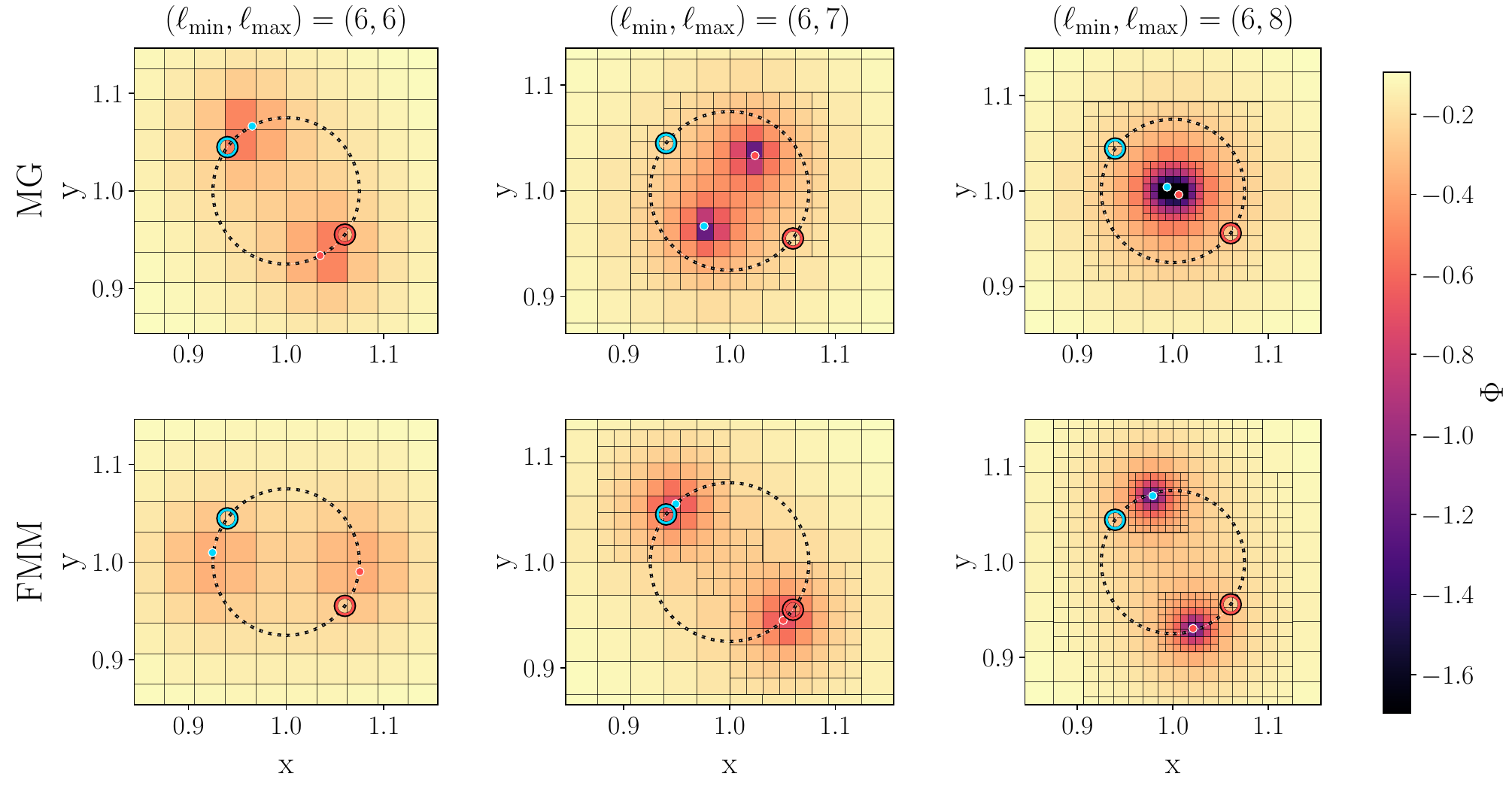}
    \caption{Same binary orbit test as Figure~\ref{fig:binary} but with $\texttt{nexpand}=1$, with less mesh smoothing. The charges are still always located at the finest level of refinement, but the volume of the refined region is much smaller than the $\texttt{nexpand}=2$ case. While FMM stays robust to the degree of mesh smoothing, MG is severely impacted and the two charges completely spiral in within three orbits.} 
    \label{fig:binary_nexpand}
\end{figure*}

\begin{figure*}
\includegraphics[width=\textwidth]{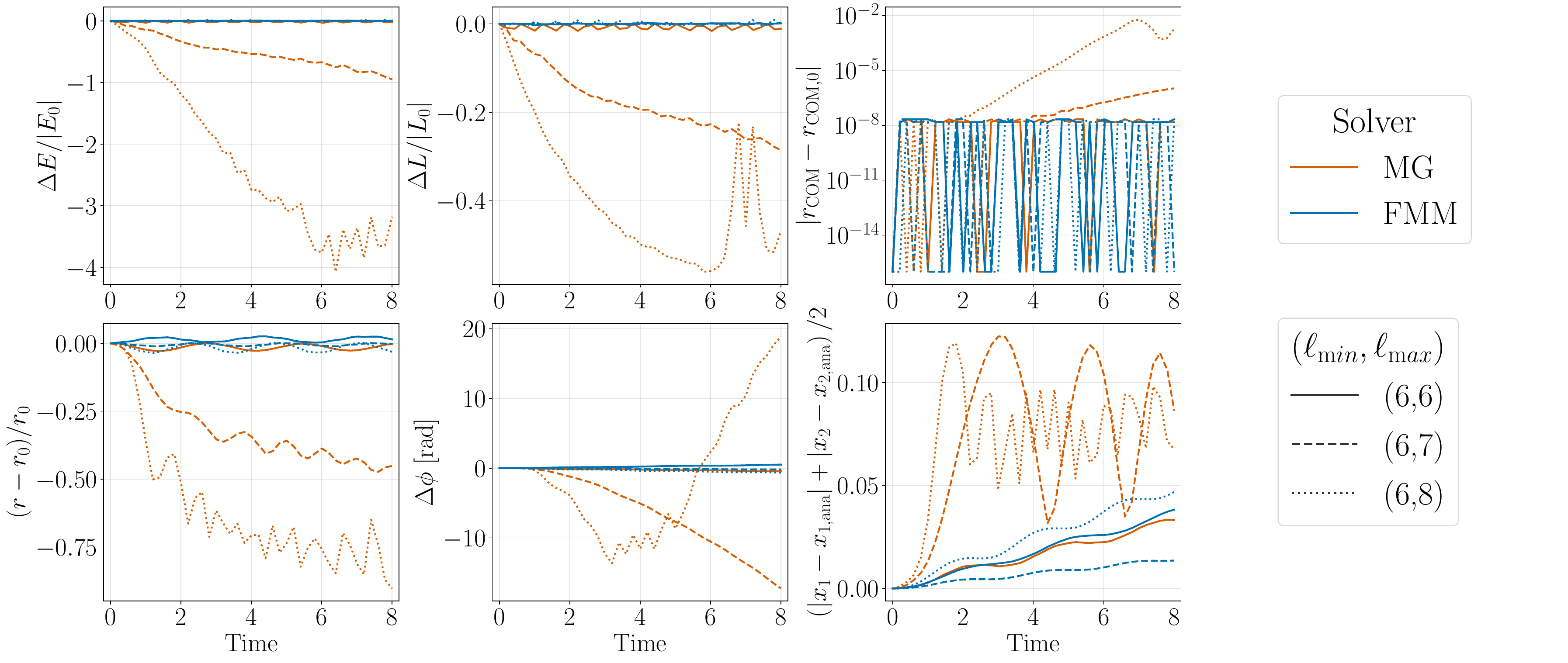}
    \caption{Diagnostics of conservation properties for the binary orbit test case in Fig.~\ref{fig:binary_nexpand} with $\texttt{nexpand}=1$.}
\label{fig:binary_nexpand_diagnostics}
\end{figure*}

In this section, we repeat the binary orbit tests shown in Figures~\ref{fig:binary} and \ref{fig:binary_diagnostics} but using $\texttt{nexpand}=1$ instead of $\texttt{nexpand}=2$. As shown in Figure~\ref{fig:binary_nexpand}, although the charges remain on the finest refinement level throughout the simulation, the refined region occupies a much smaller volume than when a higher degree of mesh expansion is applied. The unigrid configuration is unaffected, but the MG solution deteriorates rapidly even in the $(6,7)$ case, while in the $(6,8)$ case the two particles spiral completely inward within three orbits. In contrast, FMM retains a similar level of numerical stability regardless of the degree of mesh expansion. Figure~\ref{fig:binary_nexpand_diagnostics} shows the relative change in total energy ($\Delta E/E_0$), angular momentum ($\Delta L/L_0$), center-of-mass drift ($|r_{\rm COM} - r_{{\rm COM},0}|$), interparticle separation ($(r-r_0)/r_0$), phase drift ($\Delta \phi$), and the mean absolute deviation of the particle positions from the analytical solutions ($(|x_1 - x_{1,\rm ana}|+|x_2 - x_{2,\rm ana}|)/2$) for the numerical experiments run in Figure~\ref{fig:binary_nexpand}. The conservation properties of MG are substantially degraded relative to the $\texttt{nexpand}=2$ case (see Figure~\ref{fig:binary_diagnostics}), in which the orbital diagnostics exhibit only a comparatively gradual secular drift.

The lack of mesh expansion results in a less smooth mesh configuration. Consequently, the charges are more susceptible to potential-field errors arising from the inexact Dirichlet boundary conditions imposed at coarse-fine interfaces, because there is less buffer space between the charges and the interfaces. In contrast to MG, FMM is relatively robust to such non-ideal coarse-fine interface configurations because the potential field is evaluated directly using Taylor expansions rather than through imposed Dirichlet boundary conditions. Therefore, proper mesh smoothing is crucial for numerical stability in MG (for sink particles in particular), while FMM is overall stable regardless of the detailed mesh configuration.

\bsp	
\label{lastpage}
\end{document}